\documentclass[iop,useAMS,usenatbib,preprint]{emulateapj}
\usepackage{amsmath}
\usepackage{graphicx}
\usepackage{ulem}

\newcommand{\ltsimeq}{\raisebox{-0.6ex}{$\,\stackrel
        {\raisebox{-.2ex}{$\textstyle <$}}{\sim}\,$}}
\newcommand{\gtsimeq}{\raisebox{-0.6ex}{$\,\stackrel
        {\raisebox{-.2ex}{$\textstyle >$}}{\sim}\,$}}

\shorttitle{Two-Stage Fragmentation for Cluster Formation}
\shortauthors{Bailey \& Basu}

\begin{document}

\title{Two-Stage Fragmentation for Cluster Formation: Analytical Model and Observational Considerations }

\author{Nicole D. Bailey and Shantanu Basu}
\affil{Department of Physics and Astronomy, University of Western Ontario\\
 1151 Richmond Street, London, Ontario, N6A 3K7}
\email{nwityk@uwo.ca (NDB); basu@uwo.ca (SB)}

\begin{abstract}
Linear analysis of the formation of protostellar cores in planar magnetic interstellar clouds shows that molecular clouds exhibit a preferred length scale for collapse that depends on the mass-to-flux ratio and neutral-ion collision time within the cloud. We extend this linear analysis to the context of clustered star formation. By combining the results of the linear analysis with a realistic ionization profile for the cloud, we find that a molecular cloud may evolve through two fragmentation events in the evolution toward the formation of stars. Our model suggests that the initial fragmentation into clumps occurs for a transcritical cloud on parsec scales while the second fragmentation can occur for transcritical and supercritical cores on subparsec scales. Comparison of our results with several star forming regions (Perseus, Taurus, Pipe Nebula) shows support for a two-stage fragmentation model.
\end{abstract}


\section{Introduction}
\label{intro}
Molecular clouds are observed to have complex morphology existing on a wide range of scales \citep[e.g.,][]{mck07,gol08}, and exhibit numerous clumpy and filamentary structures. They are the birthplaces of stars, and the observed distribution of young stellar objects (YSOs) also yields a wide range of object densities. The YSOs are measured to be anywhere from loosely clustered to highly clustered \citep{Megeath2009}. 

Many molecular cloud maps do however reveal an interesting pattern of dark clouds within which parsec-scale clusters are being formed. The dark clouds are often themselves separated by several pc. Even the Taurus molecular cloud, often associated with so-called ``isolated'' star formation, demonstrates several dark clouds \citep{oni98} of pc-scale and masses $\sim 100$~M$_{\odot}$ that are separated by several pc, each containing embedded weak clusters of YSOs. It was suggested long ago that the dark clouds within a molecular cloud complex may be the result of some early fragmentation process \citep{Dutrey1991,Gaida1984,Schneider1979}, and subsequently the suggestion was made that the fragmentation {\it within} the dark clouds may be controlled by the local Jeans length \citep{Hartmann2002}. Based on the observed morphology of clustering within parsec-scale clouds, there is indeed a case to be made for a {\it two-stage fragmentation} process. The fragmentation within the dark clouds also shows a possible scaling with Jeans length, e.g., the typical separation between cores in the Taurus dark clouds is 0.25 pc, consistent with the Jeans length corresponding to a column density associated with the visual extinction $A_V \simeq 5$. 

The question then arises as to what controls the fragmentation process at the parsec scale, leading to the formation of individual dark clouds within a molecular cloud complex. Since these regions have nonthermal line widths, the mechanism of turbulent fragmentation, in which turbulence plays a dominant role, is often invoked \citep{Padoan1997,Gammie2003,Klessen2001}. However, turbulent fragmentation models make no unique prediction for any observed molecular cloud. The usual technique of running simulations with power spectra of the form $P(k) \propto k^{-n}$, with $n>0$, means that much of the power is contained in the largest scale of the simulation, leading to the largest structures forming on the chosen size scale of the simulation region. In this paper, we look for an explanation for large initial fragmentation scales that does not depend on the scale of the modeled system. We employ a linear analysis that assumes a uniform background state, so it does suffer from the standard ``Jeans swindle'' that is well known in the nonmagnetic analysis. We apply the analysis results using thermal line widths, without using the ansatz of replacing the isothermal sound speed with a nonthermal line width. We are in effect assuming that the fragmentation of a diffuse cloud is occurring in the dense midplane that has thermal line widths, while an extended diffuse envelope can have nonthermal line widths. Such a picture is consistent with the modeling of the propagation of turbulent motions in a stratified cloud by \citet{kud03,kud06}.

\citet{CB2006} carried out linear analysis of gravitational instability in a flattened layer that is partially ionized and threaded by a magnetic field that is initially perpendicular to the sheet. This is based on earlier work by \citet{mor91}. The results show a dramatic dependence of the preferred scale of fragmentation upon the ambient mass-to-flux ratio, if this ratio is close to the critical value. The time scale for growth of the instability also exhibits a dramatic change around the critical mass-to-flux ratio, with a step-like drop in value as the mass-to-flux ratio goes from slightly subcritical to slightly supercritical. The time scale of growth of gravitational instability in the subcritical case is commonly referred to in the literature as the ``ambipolar diffusion time''. A linear analysis of an infinite uniform medium with partial ionization and allowing for ambipolar diffusion has also been performed by \citet{lan78} and recently more comprehensively by \citet{mou11}. \citet{KM2009} have also used this linear analysis to generate a core mass function to fit the one observed in Orion. In this paper, we incorporate the new theoretical results from \citet{CB2006} regarding the fragmentation of partially ionized and magnetized gas, focusing on the fragmentation of a flattened layer, as might be an expected early configuration if the magnetic field strength is significant, or if a cloud is formed by large scale flows. Furthermore, structured objects like sheets or filaments exhibit a preferred scale of fragmentation, unlike uniform media, for which the largest scale always has the greatest growth rate. In a sheet-like situation, initial small-amplitude white noise perturbations result in the preferred scale (possessing the minimum growth time in a linear analysis) eventually growing to dominate the nonlinear evolution \citep{Basu2009b}. 

Another dramatic variation of the ambipolar diffusion time scale in a molecular 
cloud will occur because of the strong dependence of ionization fraction on column density of a molecular cloud \citep{mck89,cio95,Ruffle1998}. The ionization fraction can drop in a step-like manner as the column density of a cloud increases beyond a threshold value. The long ambipolar diffusion time scale in the UV-ionized region has been invoked as the reason for the relatively low Galactic star formation rate \citep{mck89}, and the transition to lower ionization due to cosmic rays has been invoked as a reason for lower turbulence in high column density regions \citep{Ruffle1998,mye98}. The effect of this drop in ionization upon the fragmentation process itself has not previously been investigated.
   
Based on the above findings, one can envision two dramatic events during the assemblage of a molecular cloud. If we assume that a molecular cloud is assembled from HI cloud material that is generally subcritical \citep{hei05} and ionized by UV starlight, then a cloud may resist fragmentation (due to the very long ambipolar diffusion time) until flows primarily along the magnetic field lines raise the mass-to-flux ratio to slightly above the critical value. At this point, the transcritical (but still UV ionized) cloud may form large fragments as predicted by the linear theory. As these fragments develop, they will also become more supercritical, since their evolution is partially driven by neutral-ion drift.  Once the column density in the fragments also crosses the column density threshold for the transition to cosmic ray dominated ionization, the fragmentation length and time scales will drop dramatically, and a second stage of fragmentation may be possible. This is the scenario that we investigate in this paper, by combining results of linear analysis and ionization balance.

In Section 2 we outline our model and revisit the linear analysis \citep{mor91,CB2006} with a specific focus on our chosen model parameters. In Section 3, we use the linear analysis results to discuss the feasibility of core formation by subclumps forming within a larger contracting diffuse clump. We do this by comparing numbers from the model with core and clump length scales from observed star-forming regions. A further discussion is given in Section 4 and a summary is presented in Section 5.

\section[]{Physical Model}

We consider clump/core collapse within partially ionized, isothermal, magnetic interstellar molecular clouds. These clouds are modeled as a planar sheet with infinite extent in the $x$- and $y$-directions and a local vertical half-thickness $Z$. The nonaxisymetric equations and formulations of our model have been described in detail in several papers \citep{CB2006, Basu2009a, Basu2009b}. The volume density ($\rho_{n,0}$) is calculated from the vertical pressure balance equation
\begin{equation}
\rho_{n,0} c_{s}^{2} = \frac{\pi}{2}G\sigma_{n,0}^{2} + P_{\rm ext}, 
\label{hydroequil}
\end{equation}
where $P_{\rm ext}$ is the external pressure on the sheet, $\sigma_{n,0}$ is the initial uniform column density of the sheet and $c_{s} = (k_{B}T/m_{n})^{1/2}$ is the isothermal sound speed; $k_{B}$ is the Boltzmann constant, $T$ is the temperature in Kelvins and $m_n$ is the mean mass of a neutral particle ($m_{n}= 2.33$~amu). The evolution equations also include the effect of ambipolar diffusion, which is a measure of the coupling of the neutrals with the ions, and by extension the magnetic field. This is quantified by the time scale for collisions between neutrals and ions:
\begin{equation} 
\tau_{ni} = 1.4 \left(\frac{m_i +m_{H_2}}{m_i} \right) \frac{1}{n_i\langle\sigma w\rangle_{iH_2}}.
\end{equation} 
Here, $m_{i}$ is the ion mass, and $\langle\sigma w\rangle_{iH_2}$ is the neutral-ion collision rate. The typical atomic and molecular species within a molecular cloud are singly ionized Na, Mg and HCO, which have a mass of 25 amu. Assuming collisions between H$_{2}$ and HCO$^+$, the neutral-ion collision rate is $1.69\times 10^{-9}$ cm$^{3}$ s$^{-1}$ \citep{MM1973}. These collisions transfer information about the magnetic field to the neutral particles via the ions that are bound to the field lines. The threshold for whether a region of a molecular cloud is stable or unstable to collapse is given by the normalized mass-to-flux ratio of the background reference state, 
\begin{equation}
\mu_{0} \equiv 2\pi G^{1/2}\frac{\sigma_{n,0}}{B_{\rm ref}},
\end{equation}
where $(2\pi G^{1/2})^{-1}$ is the critical mass-to-flux ratio for gravitational collapse in the adopted model \citep{NN1978,CB2006} and $B_{\rm ref}$ is the magnetic field strength of the background reference state. Regions with $\mu_{0} < 1$ are defined as subcritical, regions with $\mu_{0} > 1$ are defined to be supercritical, and regions with $\mu_{0} \simeq 1$ are transcritical. In the limit of flux freezing, the time scale for collisions ($\tau_{ni}$) is zero, implying frequent collisions between the neutral particles and ions, which therefore couples the neutrals to the magnetic field. Under these conditions, subcritical regions are supported by the magnetic field and only supercritical regions can collapse. For nonzero values of $\tau_{ni}$, a finite time scale for collisions between neutral particles and ions exists which is inversely dependent on the ion density. The variation in ion density within a cloud is linked to the type of ionizing radiation available at different depths within the cloud. This is shown in Figure 1 of \citet{Ruffle1998}. At low extinction, i.e., on the edge of the cloud, the ionization fraction is large, and declines steeply beyond a threshold column density due to shielding of the inner regions from the UV radiation that ionizes the outer layers. Ionization within these inner regions is predominantly due to cosmic rays. Ambipolar diffusion, i.e., neutral-ion slip, can lead to redistribution of mass relative to magnetic flux within a molecular cloud and cause gravitationally unstable regions to develop within subcritical regions. These regions then have the chance to collapse.

The model we use is characterized by several dimensionless free parameters. These include a dimensionless form of the initial neutral-ion collision time ($\tau_{ni,0}/t_{0}~\equiv~2\pi G\sigma_{n,0}\tau_{ni,0}/c_{s}$) and a dimensionless external pressure ($\tilde{P}_{\rm ext} \equiv 2 P_{\rm ext}/\pi G \sigma^{2}_{n,0}$), in addition to the dimensionless mass-to-flux ratio $\mu_{0}$ that was introduced earlier. We normalize column densities by $\sigma_{n,0}$, length scales by $L_{0} = c_{s}^{2}/2\pi G \sigma_{n,0}$ and time scales by $t_{0} = c_{s}/2\pi G \sigma_{n,0}$.
Based on these parameters, typical values of the units used and other derived quantities are 

\begin{eqnarray}
\nonumber\sigma_{n,0} &=& \frac{3.63\times 10^{-3}}{(1+\tilde{P}_{\rm ext})^{1/2}}\left(\frac{n_{n,0}}{10^3 \rm ~cm^{-3}}\right)^{1/2}\left(\frac{T}{10 ~\rm K}\right)^{1/2} \rm g~cm^{-2},\\ 
&&\\
c_{s} &=& 0.188\left(\frac{T}{10 ~\rm K}\right)^{1/2} \rm km~s^{-1},\\
t_{0} &=& 3.98\times 10^5\left(\frac{10^3 \rm~ cm^{-3}}{n_{n,0}}\right)^{1/2}(1 + \tilde{P}_{\rm ext})^{1/2}~\rm yr,\label{time}\\ 
\nonumber L_{0} &=& 7.48\times 10^{-2} \left(\frac{T}{10 ~\rm K}\right)^{1/2}\times\\
&&\left(\frac{10^3 \rm ~cm^{-3}}{n_{n,0}}\right)^{1/2}(1 + \tilde{P}_{\rm ext})^{1/2}~\rm pc,\label{length}\\
\nonumber\tau_{ni,0} &=& 3.74\times 10^{4} \left(\frac{T}{10~\rm K}\right)\left(\frac{0.01~\rm g~cm^{-2}}{\sigma_{n,0}}\right)^{2}\times\\
&&\left(\frac{10^{-7}}{\chi_{i,0}}\right)(1+\tilde{P}_{\rm ext})^{-1}~ \rm yr,\label{eqn:tauni}
\end{eqnarray}
where $n_{n,0}$ is the initial neutral number density and $\chi_{i,0}$ the ionization fraction. Note that $\tau_{ni,0}$ is inversely proportional to $\chi_{i,0}$ for a fixed $\sigma_{n,0}$. For our analysis, we assume a dimensionless external pressure $\tilde{P}_{\rm ext} = 0.1$ and a temperature $T = 10$ K.

\begin{figure*}[!h]
\includegraphics[width=0.5\textwidth]{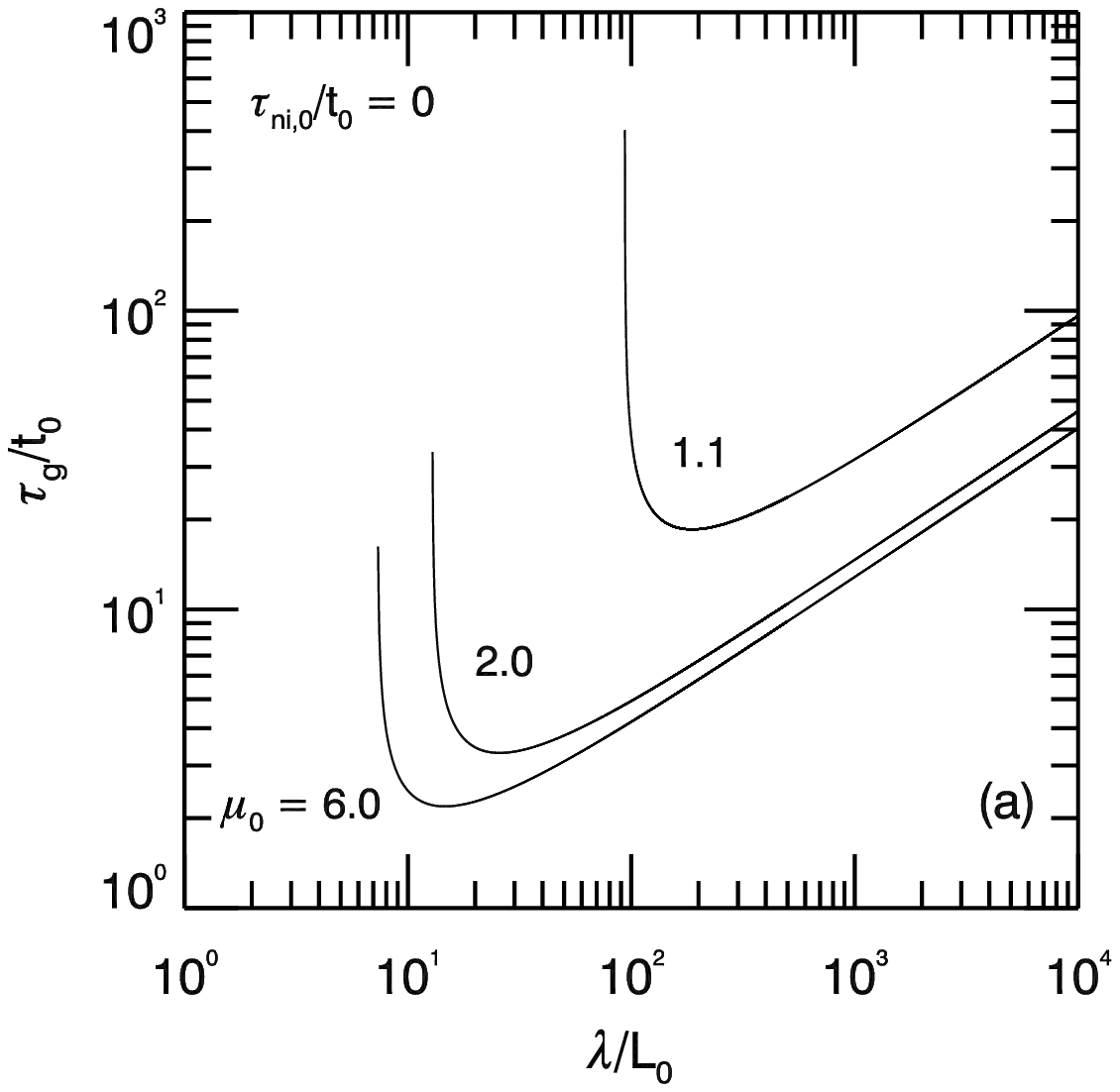}
\includegraphics[width=0.5\textwidth]{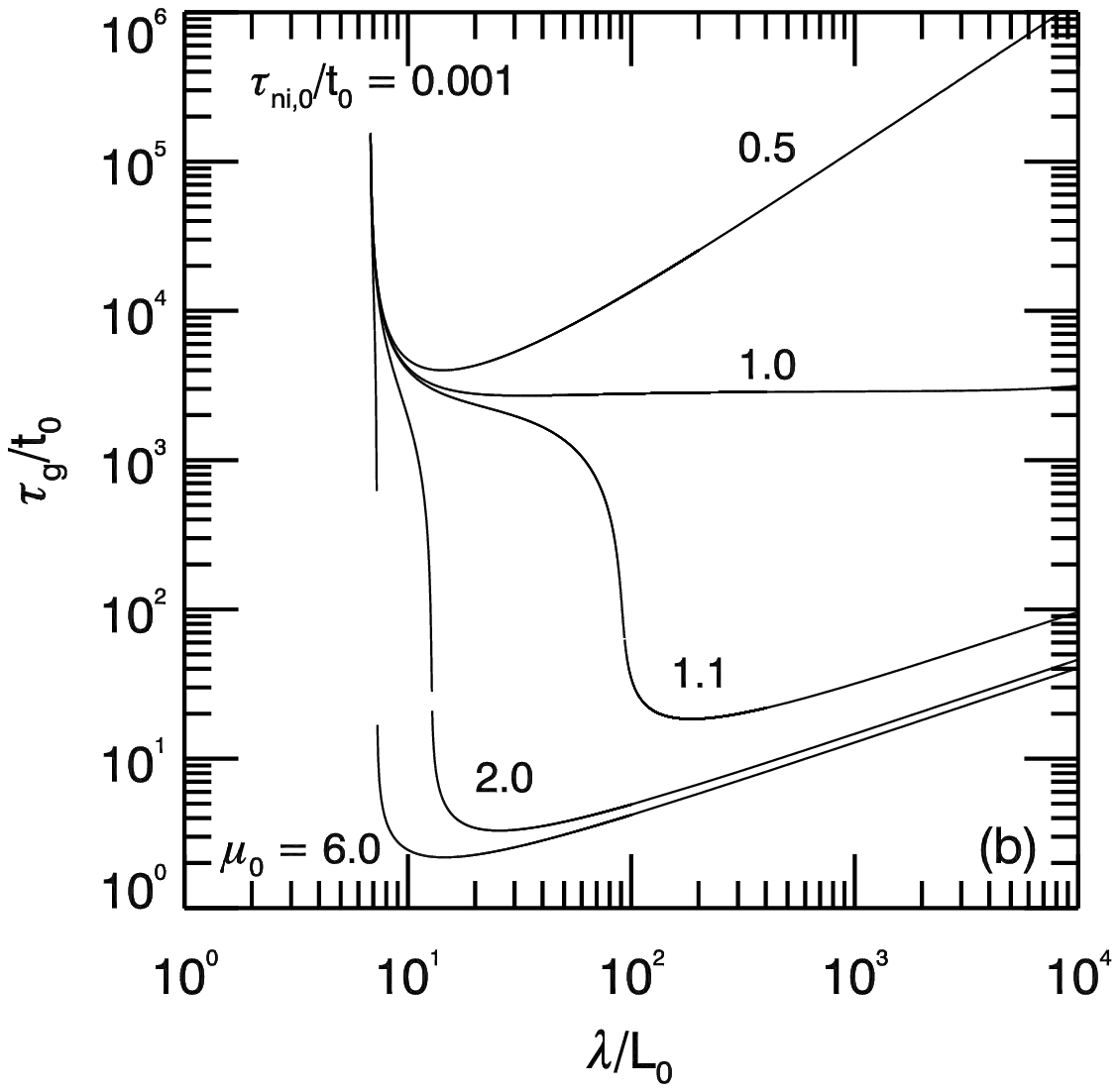}\\
\includegraphics[width=0.5\textwidth]{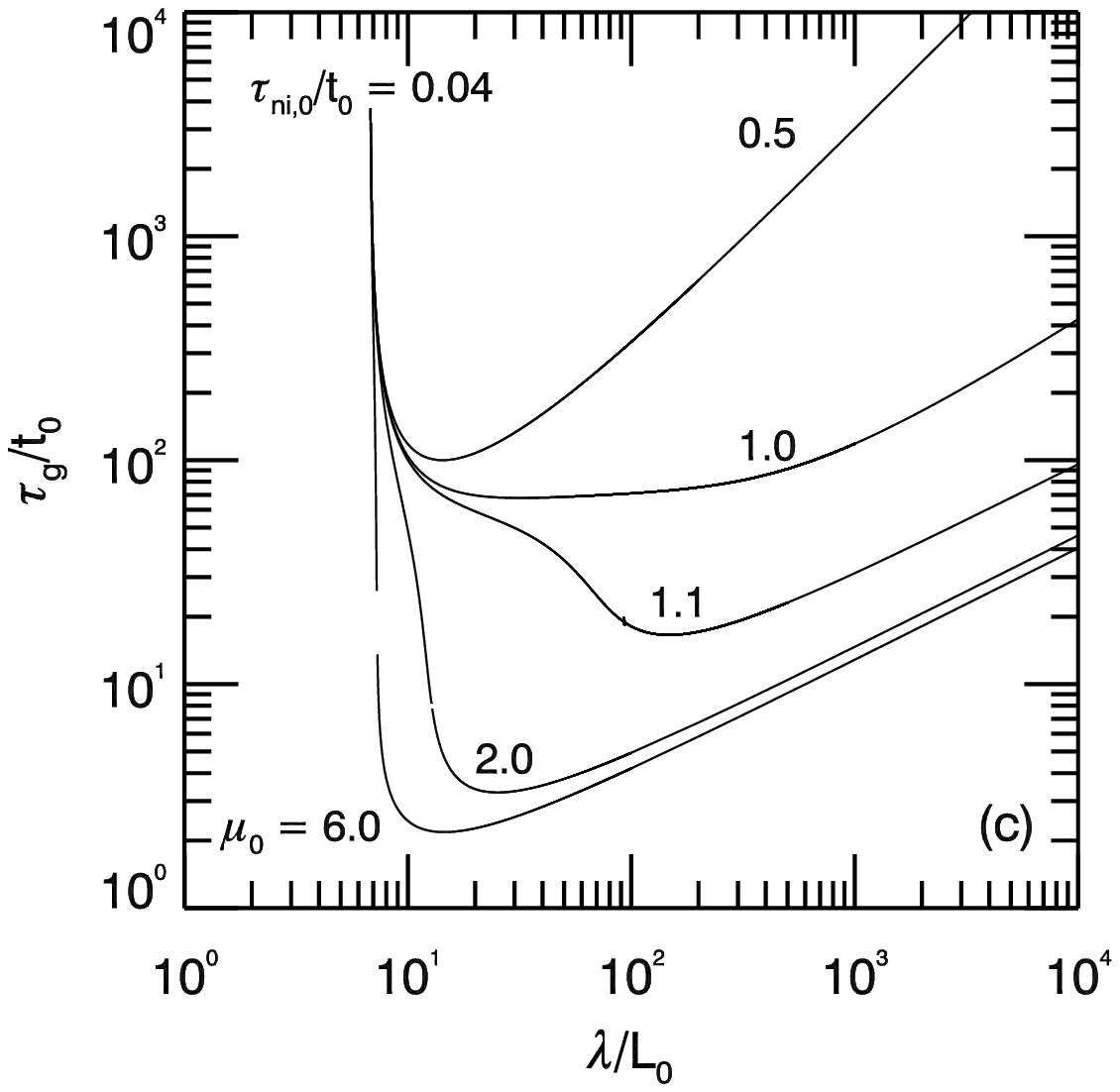}
\includegraphics[width=0.5\textwidth]{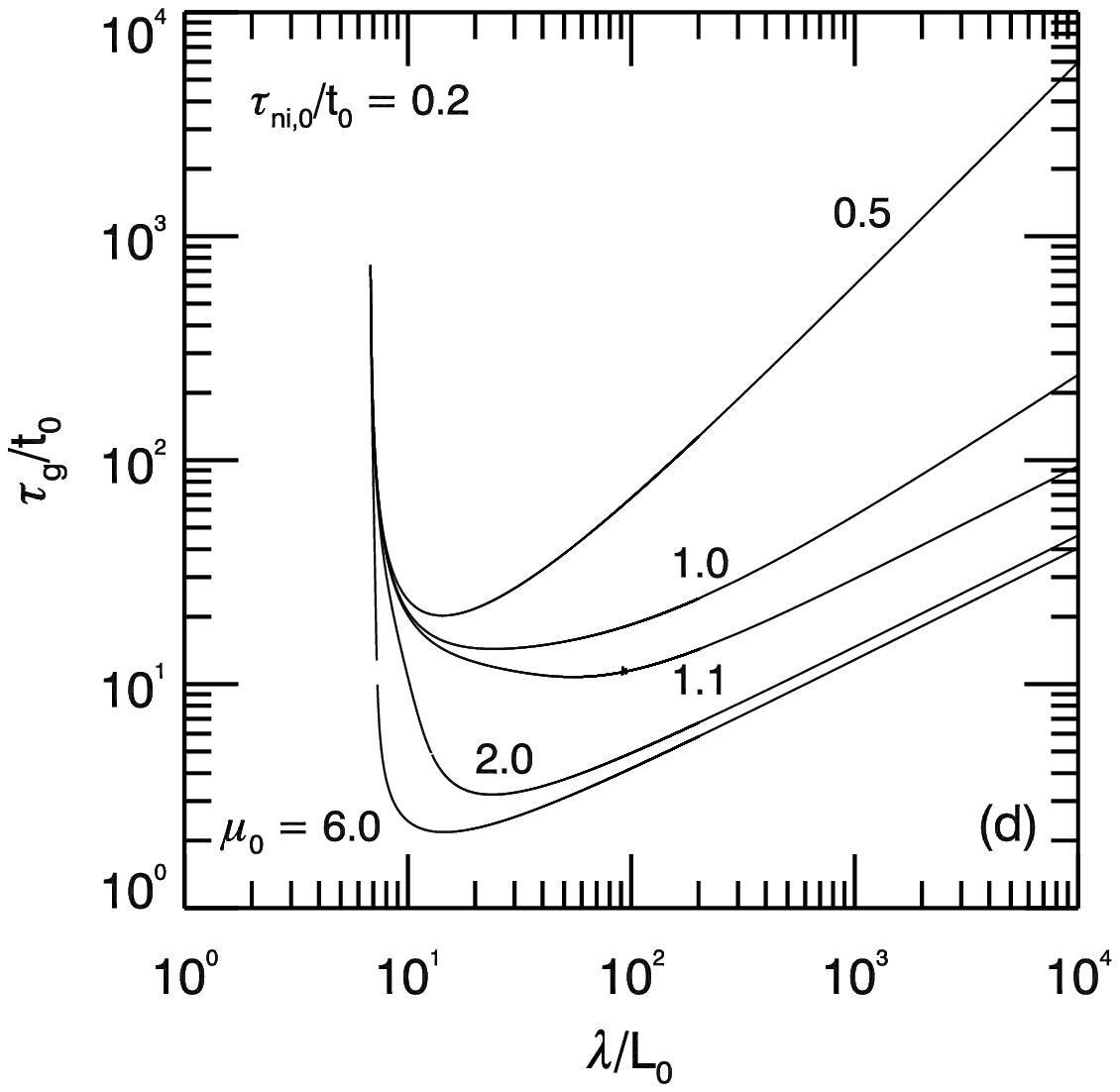}
\caption{Growth time of gravitationally unstable mode ($\tau_{g}/t_{0}$) as a function of wavelength ($\lambda/L_{0}$) for models (a) $\tau_{ni,0}/t_{0}$ = 0, (b) $\tau_{ni,0}/t_{0}$ = 0.001, (c) $\tau_{ni,0}/t_{0}$ = 0.04, and (d) $\tau_{ni,0}/t_{0}$ = 0.2. Each panel shows time scale curves for models with mass-to-flux ratios $\mu_{0}~=~0.5,~1.0,~1.1,~2.0,$~and~$6.0$ (labeled). }
\label{fig:taulambda}
\end{figure*}
\begin{figure*}
\includegraphics[width=0.5\textwidth]{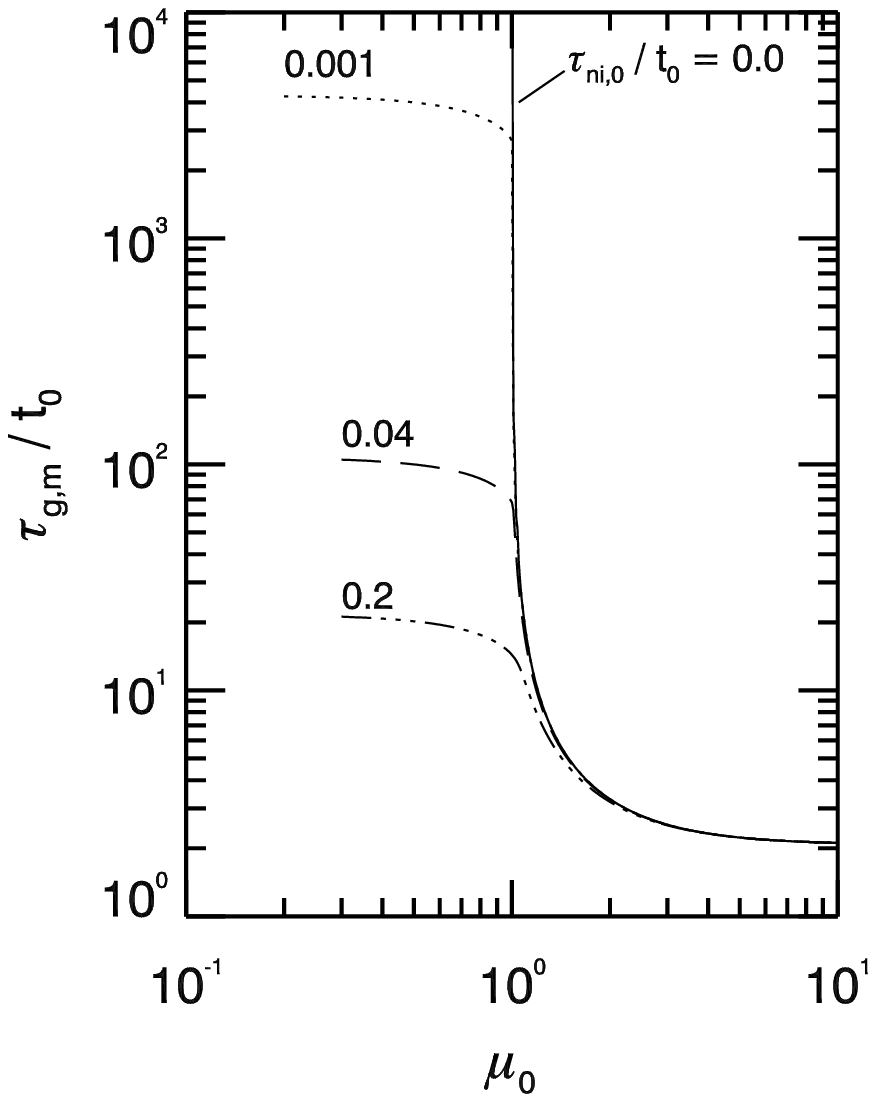}
\includegraphics[width=0.5\textwidth]{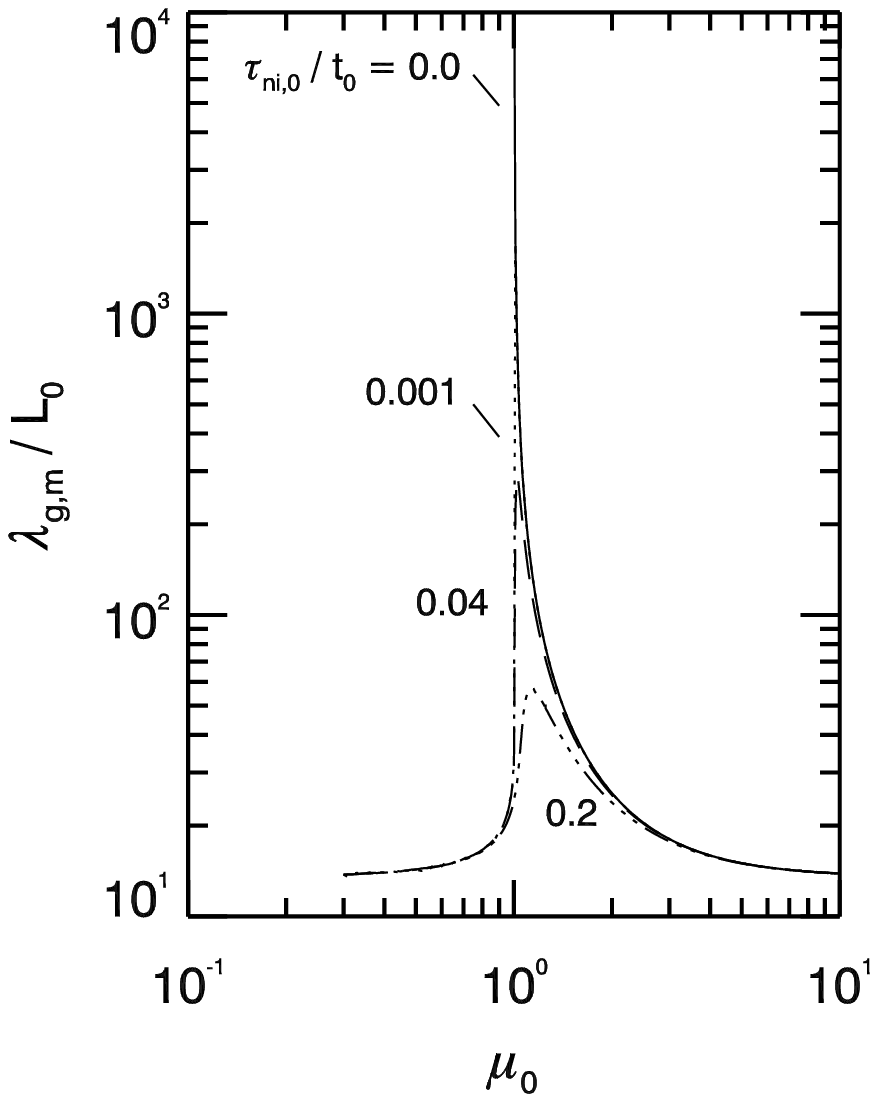}
\caption{Left: Minimum growth time of gravitationally unstable mode ($\tau_{g,m}/t_{0}$) as a function of mass-to-flux ratio ($\mu_{0}$).  Right: Length scale of most unstable mode ($\lambda_{g,m}/L_{0}$) as a function of the mass-to-flux ratio.}
\label{fig:vsmu}
\end{figure*}
\begin{figure*}
\includegraphics[width=0.5\textwidth]{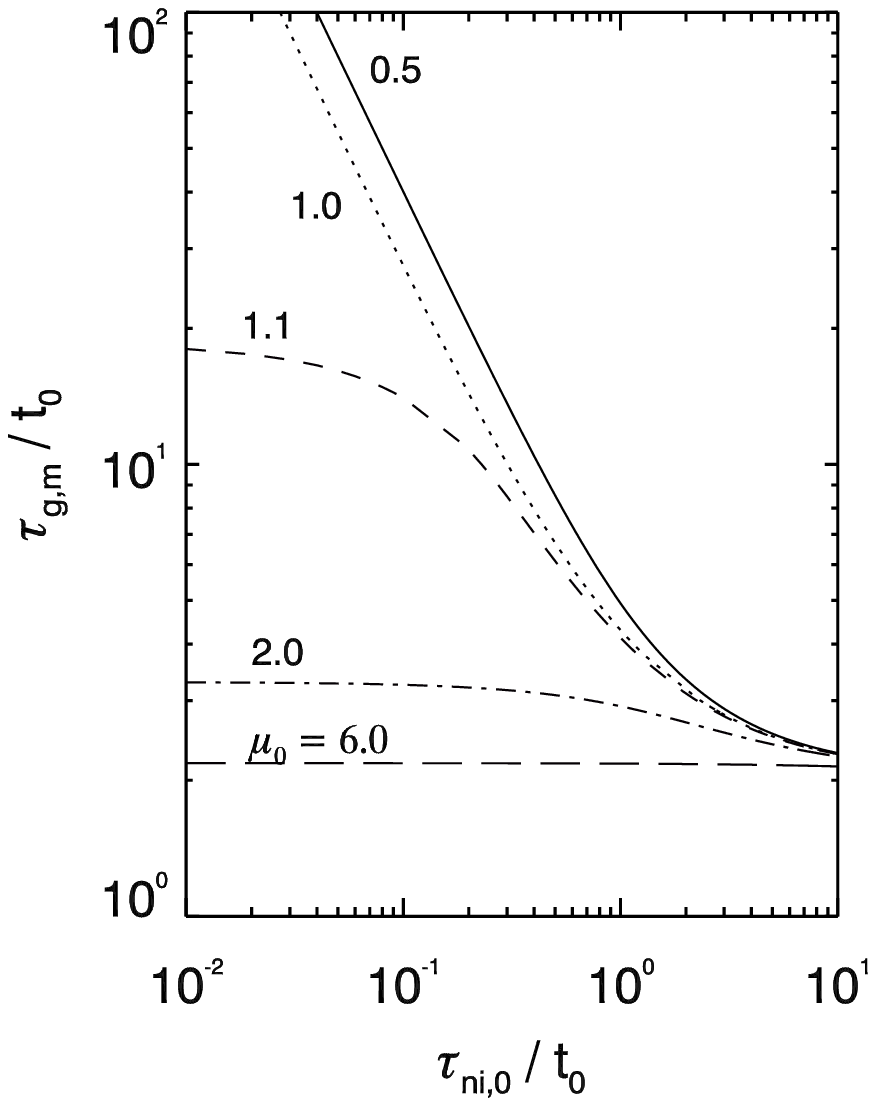}
\includegraphics[width=0.5\textwidth]{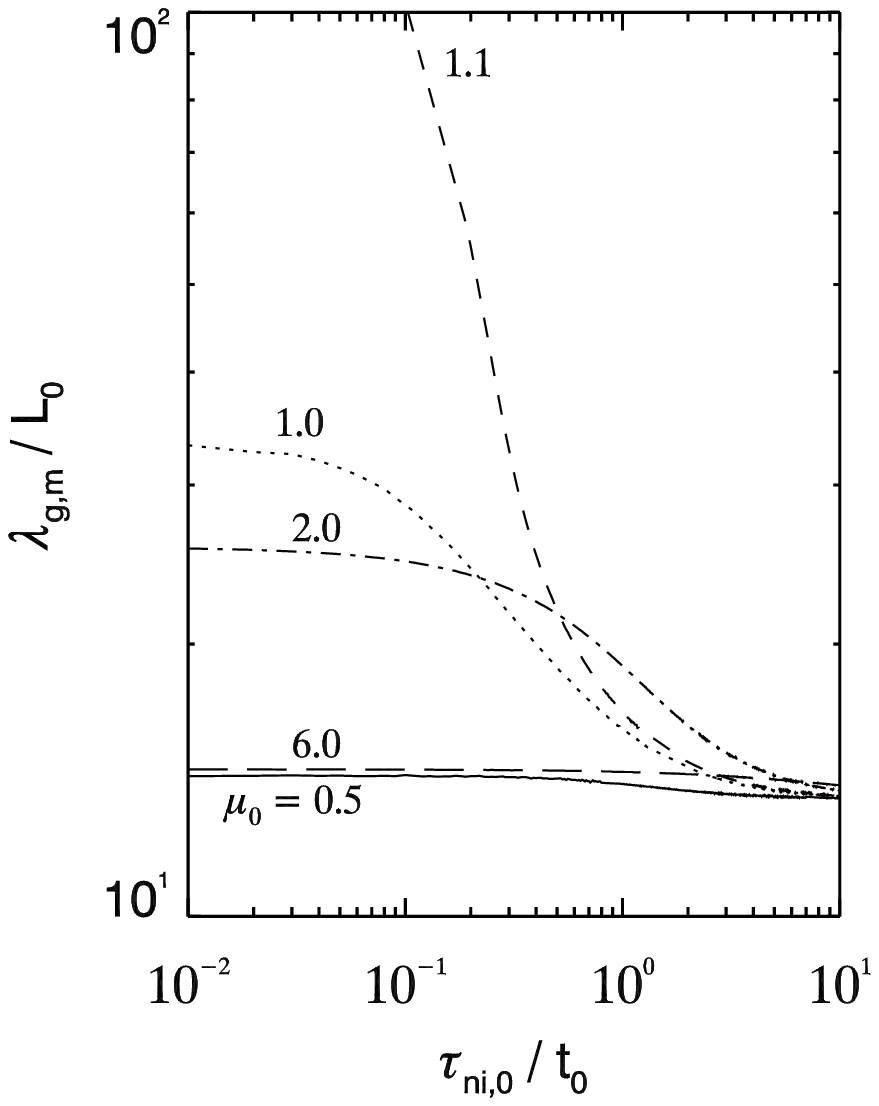}
\caption{Left: Minimum growth time ($\tau_{g,m}/t_{0}$) as a function of the dimensionless initial neutral-ion collision time (${\tau}_{ni,0}/t_{0}$). Right: Most unstable length scale ($\lambda_{g,m}/L_{0}$) as a function of the dimensionless initial neutral-ion collision time.}
\label{fig:vstauni}
\end{figure*}

\subsection{Linear Analysis}

Assuming the unperturbed zero-order or background state of the model is static and uniform, the magnetohydrodynamic equations for a thin sheet \citep[see][]{Basu2009a} can be linearized to first order for any physical quantity via
\begin{eqnarray}
\nonumber f(x,y,t) &=& f_{0} + \delta f\\
&=& f_{0} + \delta f_{a} e^{i(k_{x}x+k_{y}y-\omega t)},
\end{eqnarray} 
where $f_{0}$ is the unperturbed background state, $\delta f$ is the perturbation, $\delta f_{a}$ is the amplitude of the perturbation, $k_{x}$, $k_{y}$, and $k$ are the $x$-, $y$-, and $z$- wavenumbers respectively such that $k^{2}\equiv k_{x}^{2} + k_{y}^{2}$  and $\omega$ is the complex angular frequency. The perturbations are necessarily small such that $|\delta f_{a}| \ll f_{0}$. With this assumed perturbation, $\partial/\partial t \rightarrow -i\omega$, $\partial/\partial x \rightarrow ik_{x}$, and $\partial/\partial y \rightarrow ik_{y}$. Applying the perturbation to the dimensional equations for a model cloud (Equations (2) - (4) of \citet{Basu2009a}), and retaining only the first order terms, one obtains, in the manner of \citet{CB2006},
\begin{eqnarray}
\omega\delta\sigma_{n,0} &=& k_x c_s\delta v_{ n,x} + k_y c_s\delta v_{n,y},\\
\nonumber\omega c_s\delta v_{n,x} &=& \frac{k_x}{k}(C^2_{\rm eff,0}k - 2\pi G \sigma_{n,0})\delta\sigma_{n,0} \\
                                  &+& \frac{k_x}{k}(2\pi G\sigma_{n,0}\mu^{-1}_0 + kV_{A,0}^{2}\mu_{0})\delta B_{z,\rm eq},\\
\nonumber\omega c_s \delta v_{n,y} &=& \frac{k_y}{k}(C^2_{\rm eff,0}k - 2\pi G \sigma_{n,0})\delta\sigma_{n,0} \\
                                   &+& \frac{k_y}{k}(2\pi G\sigma_{n,0}\mu^{-1}_0 + kV_{A,0}^{2}\mu_{0})\delta B_{z,\rm eq},\\
\nonumber\omega\delta B_{z,\rm eq} &=& \frac{1}{\mu_{0}}k_x c_s\delta v_{n,x} + \frac{1}{\mu_{0}}k_y c_s\delta v_{n,y}\\
\nonumber & & -i\tau_{ni,0}(2\pi G \sigma_{n,0}\mu_{0}^{-2}k + k^2V_{A,0}^2)\delta B_{z,\rm eq},  \\
&&                                        
\end{eqnarray}
where $v_{n,x}$ and $v_{n,y}$ are the velocities of the neutral particles in the $x$- and $y$- directions respectively, $B_{z,\rm eq}$ is the magnetic field strength at the midplane of the sheet, $C_{\rm eff,0}$ is the local effective sound speed, where
\begin{equation}
C_{\rm eff,0}^{2} = \frac{\pi}{2}G\sigma_{n,0}^{2}\frac{[3P_{\rm ext}+(\pi/2)G\sigma_{n,0}^{2}]}{[P_{\rm ext}+(\pi/2)G\sigma_{n,0}^{2}]^{2}}c_{s}^{2},
\end{equation}
$V_{A,0}$ is the Alfv\'en speed which is related to the mass-to-flux ratio ($\mu_{0}$) via
\begin{equation}
V_{A,0}^{2} \equiv \frac{B_{\rm ref}^{2}}{4\pi\rho_{n,0}} = 2\pi G\sigma_{n,0}\mu_{0}^{-2}Z_{0},
\end{equation}
and $Z_{0}$ is the initial half-thickness of the sheet.

A mode is unstable if the imaginary part of the complex frequency $\omega_{\rm IM} > 0$. The growth time of such an instability is  $\tau_{g} = 1/\omega_{\rm IM}$. The dispersion relation is found to be
\begin{eqnarray}
\nonumber(\omega&+&i\theta)(\omega^{2} - C_{\rm eff,0}^2 k^2 + 2\pi G \sigma_{n,0}k) \\
                &=& \omega(2\pi G\sigma_{n,0}k\mu^{-2}_{0} + k^2V_{A,0}^2),
\end{eqnarray}
where
\begin{equation}
\theta = \tau_{ni,0}(2\pi G\sigma_{n,0}k\mu^{-2}_{0} + k^2V_{A,0}^2).
\end{equation}
In the limit of flux freezing, $\tau_{ni,0} \rightarrow 0$, we obtain the reduced dispersion relation,
\begin{equation}
\omega^{2} + 2\pi G \sigma_{n,0}k(1-\mu^{-2}_{0})- k^{2}(C_{\rm eff,0}^2+V_{A,0}^2) = 0
\end{equation}
The gravitationally unstable mode corresponds to one of the roots of $\omega^2 < 0$ and occurs for $\mu_{0} > 1$. The growth time for this mode can be written as
\begin{equation}
\tau_{g} = \frac{\lambda}{2\pi[G\sigma_{n,0}(1-\mu_0^{-2})(\lambda - \lambda_{MS})]^{1/2}},
\label{eqn:taug}
\end{equation}
for $\lambda \geq \lambda_{MS}$, where
\begin{equation}
\lambda_{MS} =\frac{{C}_{\rm eff,0}^2 + V_{A,0}^2}{G\sigma_{n,0}(1 - \mu_{0}^{-2})}.
\label{lammu}
\end{equation}

With the inclusion of ambipolar diffusion, the gravitationally unstable mode still corresponds to one of the roots of the full dispersion relation; however since it is a cubic function, the expression corresponding to the roots cannot be written down as simply as Equation~(\ref{eqn:taug}) and therefore are evaluated numerically. 

Figure~\ref{fig:taulambda} shows the growth time $\tau_{g}/t_{0}$ as a function of the wavelength for four cases of the neutral-ion collision time $\tau_{ni,0}/t_{0}$. Each panel shows the dependence for several labeled values of $\mu_{0}$. Note that these figures are presented in normalized form. These different cases were chosen to represent various areas within a molecular cloud: diffuse regions with high ionization fractions ($\tau_{ni,0}/t_{0} = 0.001$), dense core forming regions with low ionization fractions ($\tau_{ni,0}/t_{0} = 0.2$) and an intermediate region that is somewhat between the two extremes ($\tau_{ni,0}/t_{0}= 0.04$). The flux-frozen case serves as a reference point. The value of $\tau_{ni,0}/t_{0}= 0.001$ was chosen to correspond to a visual extinction $A_{v} = 1$ in our adopted ionization model for the cloud (see \S~\ref{ctoc}, Figure~\ref{xivssigma}). Specifically, this value corresponds to a column density $\sigma_{n}= 0.004 \rm~g~cm^{-2}$.

For the flux-frozen case, note the dependence on the value of $\mu_{0}$; there are no unstable, gravitationally collapsing modes for $\mu_{0} < 1$. This is consistent with the discussion earlier that for a flux-frozen cloud, only initially supercritical clouds can collapse. For the other three cases however, the addition of ambipolar diffusion allows for unstable, gravitationally collapsing modes to exist for regions with $\mu_{0} < 1$. For all four plots and all cases of $\mu_{0}$ shown, note that each curve has a distinct minima. This minima represents the shortest growth time (fastest growthrate) and corresponding preferred length scale for instability. We find the location of these minima for a series of values of $\mu_{0}$ for fixed values of $\tau_{ni,0}/t_{0}$. The left panel of Figure \ref{fig:vsmu} shows the minimum growth time of the gravitationally unstable mode ($\tau_{g,m}/t_0$) as a function of the critical mass-to-flux ratio ($\mu_{0}$). In the limit of flux-freezing, the curve shows that for the supercritical regime ($\mu_{0} > 1$), the growth time for instability is short, essentially the dynamical time $2t_{o} \approx Z_{0}/c_{s}$. As the mass-to-flux ratio approaches the critical value ($\mu_0 = 1$) the growth time scale for instability becomes infinitely long. Consistent with the discussion above, this implies that in the absence of ambipolar diffusion, it would take an infinite amount of time for a region with a critical or subcritical mass-to-flux ratio to collapse into a core. As the mass-to-flux ratio approaches infinity, this implies negligible magnetic support. In this regime, the growth time $\tau_{g,T}$ is dependent on the critical thermal length scale ($\lambda_{T} \equiv G\sigma_{n,0}C_{\rm eff,0}^2$) as follows,
\begin{equation}
\tau_{g,T} = \frac{\lambda}{2\pi[G\sigma_{n,0}(\lambda - \lambda_{T})]^{1/2}}.
\end{equation}
The minimum growth time for the unstable mode occurs at $\lambda_{T,m} = 2\lambda_{T}$. With the addition of ambipolar diffusion, the curves show that the subcritical regime has a finite growth time scale for instabilities. The magnitude of this time scale depends on the degree of ambipolar diffusion, i.e., the time scale for neutral-ion collisions. For short neutral-ion collision times ($\tau_{ni,0}/t_{0} < 0.2$), the time scale for collapse of a subcritical region is 10~-~50 times longer than that of a supercritical region; frequent collisions between the neutrals and ions will slow the redistribution of matter across the field lines. This is the origin of the often quoted result that the ambipolar diffusion time is $\approx$~10 times the free-fall time. As the neutral-ion collision time increases (i.e., $\tau_{ni,0}/t_{0} > 0.2$), the time between collisions increases to a point that neutrals rarely/never collide with an ion, and therefore have no knowledge of the magnetic field present in the molecular cloud. In this case, the time scale for collapse of the subcritical regions approaches that of thermal collapse ($\tau_{g,T}$).

The right panel of Figure~\ref{fig:vsmu} shows the wavelength with minimum growth time, $\lambda_{g,m}/L_{0}$, as a function of $\mu_{0}$ \citep[see also][]{mor91,CB2006}. There are several features of this graph that are worth noting here. In the flux-frozen case, as the mass-to-flux ratio approaches the critical value ($\mu_{0} = 1$) the wavelength with the minimum growth time approaches infinity. For a non-zero neutral-ion collision time, the wavelength with the minimum growth time becomes finite for mass-to-flux ratios less than or equal to the critical mass-to-flux ratio. This results in a peak in the region of $\mu_{0} \sim 1$. Note that the location of this peak changes, becoming slightly more supercritical as the neutral-ion collision time increases.

An alternate way to look at this data is to plot the minimum growth time and length scales as a function of the neutral ion-collision time for a constant mass-to-flux ratio. This is shown in Figure~\ref{fig:vstauni}. The panels show the minimum time scale for collapse ($\tau_{g,m}/t_{0}$, left) and corresponding wavelength ($\lambda_{g,m}/L_{0}$, right) as a function of the dimensionless neutral-ion collision time ($\tau_{ni,0}/t_{0}$) for the five cases of the initial mass-to-flux ratio ($\mu_{0}$) displayed in Figure~\ref{fig:taulambda}. Focusing on the left panel first, for $\mu_{0} \le 1.0$, the growth time decreases linearly until the neutral-ion collision time reaches unity, after which they asymptote to the thermal collapse time. Conversely, for $\mu_{0} = 2.0$, there is a plateau for highly ionized regions where the collapse time is longer while for low ionization fractions the collapse time is again the thermal collapse time. For large enough values of $\mu_{0}$, the gravitational effects outweigh the magnetic effects and the cloud will just collapse on the thermal time scale. Moving to the right panel, the general trends shown by all curves indicate that the wavelength with the minimum growth time is a maximum for short neutral-ion collision times and a minimum for long neutral-ion collision times. For regions with high ionization fractions ($\tau_{ni,0}/t_{0}\sim 0.1$), as $\mu_{0}$ increases, the wavelength increases from the thermal wavelength ($\lambda_{T}$) to a maximum at $\mu_{0} = 1.1$, and then decreases back to the thermal wavelength as $\mu_{0}$ rises further (see Figure~\ref{fig:vsmu}). On the other extreme, for low ionization fractions, the variation between the different mass-to-flux ratios is almost indistinguishable since the neutrals are poorly coupled to the ions and the magnetic field. Note however, that for intermediate values of neutral-ion collision times, $\tau_{ni,0}/t_{0} \gtsimeq 1.0$, the preferred wavelength for $\mu_{0} = 2.0$ becomes larger than those for $\mu_{0} = 1.0$ and $\mu_{0} = 1.1$. This is a result of the complex interplay between ambipolar diffusion and field-line dragging in this hybrid regime of moderate coupling. 

\section{Clumps, Cores and the Effect of $\mu_{0}$}
\label{ctoc}

Observational studies of star forming regions have shown that there are several different extinction thresholds which define the state of a molecular cloud. \citet{Ruffle1998} identify a threshold of $A_{v} \sim 3$ whereby above this value clumps within the Rosette Molecular cloud are believed to have embedded stars. Studies of the Ophiuchus and Perseus clouds by \citet{Johnstone2004} and \citet{Kirk2006}, respectively, suggest that there is a core formation threshold of $A_{v} \sim$ 5 and a star formation threshold of $A_{v} \sim 7-8$ magnitudes \citep[see also][]{oni98,Froebrich2010}. 

The ionization state of a molecular cloud depends on its density and consequently the visual extinction. As demonstrated by \citet{Ruffle1998}, the ionization profile of a molecular cloud is one in which the outer layers are highly ionized due to photoionization by background ultraviolet (UV) starlight, while the inner layers are shielded from the UV photons and ionized by cosmic rays. This picture assumes a quiescent molecular cloud which has no evidence of star formation in its past; radiation from previous generations of stars can greatly complicate this picture. 

In addition to the extinction thresholds, the linear analysis discussed above showed that there are correlations between the mass-to-flux ratio of the region and the expected time and length scales for collapse. With regards to time scales, subcritical regions evolve more slowly than supercritical regions. Conversely, with regards to length scales, highly supercritical or subcritical regions can collapse on smaller length scales than transcritical regions (i.e., those with $\mu_{0}\sim1$). The following subsections outline our method and results for our two-stage fragmentation model.

\subsection{Model and Calculations}
\label{modcalc}
In their investigation, \citet{Ruffle1998}, modeled the chemical evolution of a collapsing clump using a simplified model. Here, we have constructed a model ionization profile for the cloud similar to those shown in Figure 1 of \citet{Ruffle1998}. Unlike their ionization profiles however, ours takes into account both the UV and cosmic ray (CR) ionization regions. The UV region is described by a step function of the form 
\begin{eqnarray}
\nonumber\log \chi_i =\log \chi_{i,0} + &0.5&(\log \chi_{i,c} - \log \chi_{i,0})\times\\
&&\left(1 + \tanh\frac{A_{v}-A_{v,crit}}{A_{v,d}}\right),
\end{eqnarray}
where $\chi_{i}$ is the ionization fraction, and $\chi_{i,0}$, $\chi_{i,c}$, $A_{v,crit}$ and $A_{v,d}$ are the step function parameters, namely the maximum ionization, minimum ionization, location and width of the step respectively. The cosmic ray regime is derived from the well known power-law expression for the ion density ($n_{i}$) as a function of neutral density ($n_{n}$), namely 
\begin{equation}
n_i = X n_{n}^{1/2},
\label{ni} 
\end{equation}
where $X = 10^{-5} \rm cm^{-3/2}$ \citep{Elmegreen1979,UN1980,Tielens2005}. As shown by \citet[][Figure 11]{Caselli1998}, clear correlations between $n_{i}$ and $n_{n}$ are difficult to establish observationally. Detailed calculations using grain chemistry \citep{Ciolek1994} show that the exponent in the above relation is not a fixed value. However, a value of 1/2 is a reasonable theoretical average for the relatively low density regime considered here. We combine Equation~(\ref{ni}) with vertical hydrostatic equilibrium (Equation~\ref{hydroequil}). A smooth transition between these two regimes is chosen to occur at a visual extinction of $A_{v,CR} = 6.365$. The full profile is then described by
\begin{equation} 
\log \chi_{i} = \left\{ \begin{array}{ll}
\log \chi_{i,0} + 0.5(\log \chi_{i,c} - \log \chi_{i,0})\times& \\
~~~~~~\left(1 + \tanh\frac{A_{v}-A_{v,crit}}{A_{v,d}}\right) & \mbox{$A_{v}\le A_{v,CR}$} \\
\log[ 1.148\times10^{-7}(1 +\tilde{P}_{\rm ext})^{1/2}\times  &\\
~~~~~~\left(\frac{T}{10~\rm K}\right)^{1/2}\left(\frac{2.75~ \rm mag}{A_{v}}\right)] & \mbox{$A_{v} > A_{v,CR}$ },
\end{array}
\right.
\label{eqn:ionmodel}
\end{equation}
where the values of the step function parameters are $\log \chi_{i,0} = -4$, $\log \chi_{i,c} = -7.3203$, $A_{v,crit} = 4.0$ and $A_{v,d}=1.05$. Four decimal place accuracy is used for some of the profile parameters to ensure a smooth transition, rather than for empirical reasons. The midpoint of the drop in ionization ($A_{v,crit}$) has been adopted to be closer to $A_v \sim 5$ for comparison to observations in the Perseus region (see Section \ref{comptoobs}). All numbers above are to be taken as typical values, and our intention in this paper is to outline a general scenario rather than performing a detailed parameter search.

\begin{center}
\begin{figure}
\centering
\includegraphics[width=0.5\textwidth,angle=0]{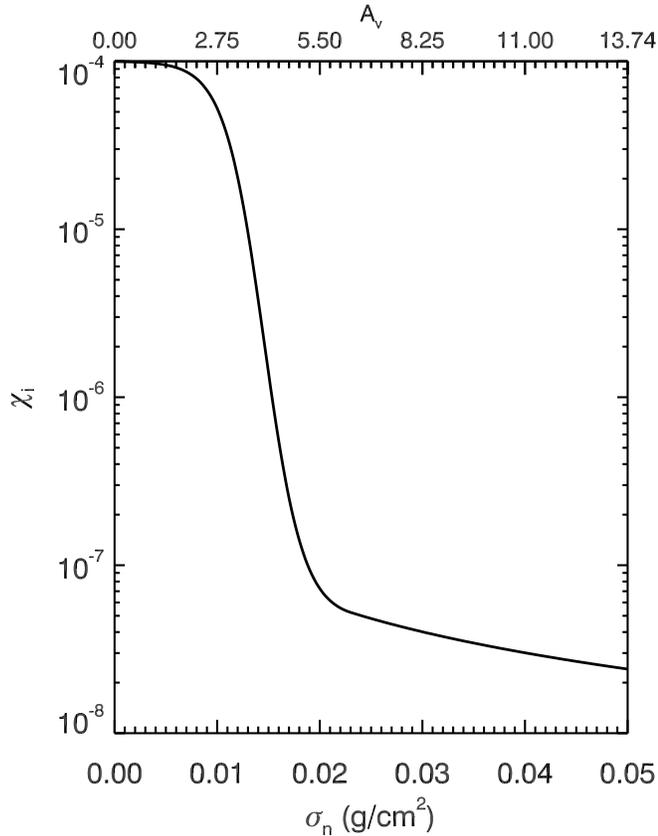}
\caption{Model ionization curve for a molecular cloud. Ionization fraction ($\chi_{i}$) is plotted against neutral column density ($\sigma_{n}$) and corresponding visual extinction ($A_{v}$) based on Equation~(\ref{av2sigma}). The curve is a composite which captures both the photoionized region ($A_{v} \leq 6.365$ mag; $\sigma_{n} < 0.0232 \rm~g~cm^{-2}$) and the cosmic-ray-ionized regions ($A_{v} > 6.365$ mag; $\sigma_{n} > 0.0232\rm~g~cm^{-2}$).}
\label{xivssigma}
\end{figure}
\end{center}

The above profile can also be expressed in terms of the column density instead of the visual extinction. Following the prescription of \citet{Pineda2010}, the conversion from visual extinction to column density can be achieved by combining the ratio of H$_{2}$ column density to color excess \citep{Bohlin1978} with the total selective extinction \citep{Whittet2003} to yield a conversion factor of $N(H_{2}) = 9.35\times 10^{20} (A_{v}/1~\rm mag)~\rm cm^{-2}$. Although this conversion is specifically for H$_{2}$, the abundance ratios of other molecules present in the cloud are so much smaller (i.e., the abundance ratio of CO to H$_{2}$ is on the order of 10$^{-4}$) that they do not contribute significantly to the overall number density. Therefore we assume that the H$_{2}$ number density is representative of all species. Assuming a mean molecular weight of 2.33 amu, this translates into a mass column density conversion of the form
\begin{equation} 
\sigma_{n} =3.638\times 10^{-3} (A_{v}/\rm mag)~\rm g~cm^{-2}.
\label{av2sigma}
\end{equation}
Figure~\ref{xivssigma} shows the resulting profile for $\chi_{i}$ as a function of $\sigma_{n}$. As shown, the adopted model is one which has a steep step function as the ionization fraction drops off at $A_{v} \sim 3$. However, instead of leveling off at the bottom as the step function would normally, it continues to decrease as dictated by the cosmic ray relation. From this model ionization structure of the cloud, the neutral-ion collision time as a function of column density and ionization fraction (Equation~\ref{eqn:tauni}) can be derived.

\begin{center}
\begin{figure}
\includegraphics[width=0.35\textwidth,angle=-90]{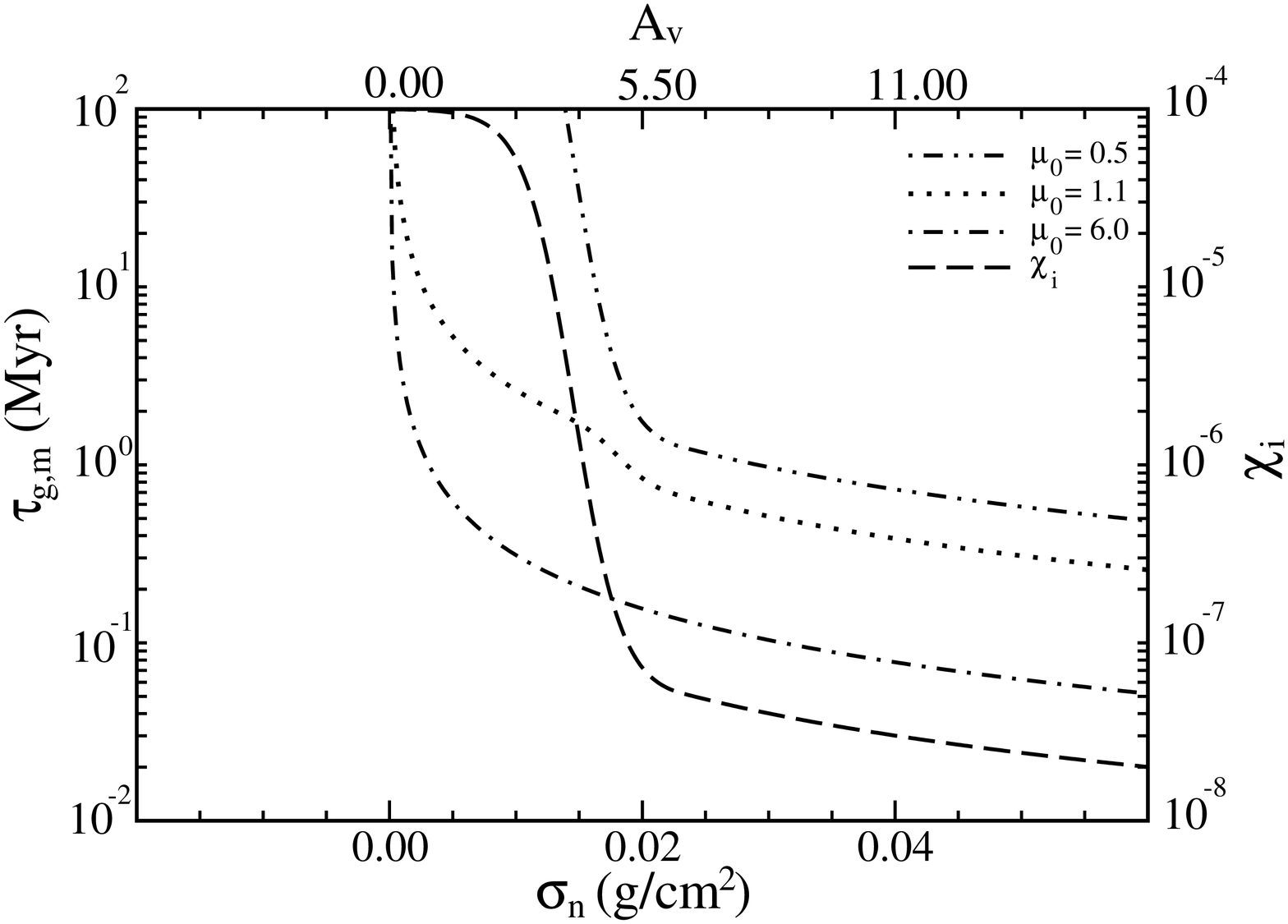}\\
\includegraphics[width=0.35\textwidth,angle=-90]{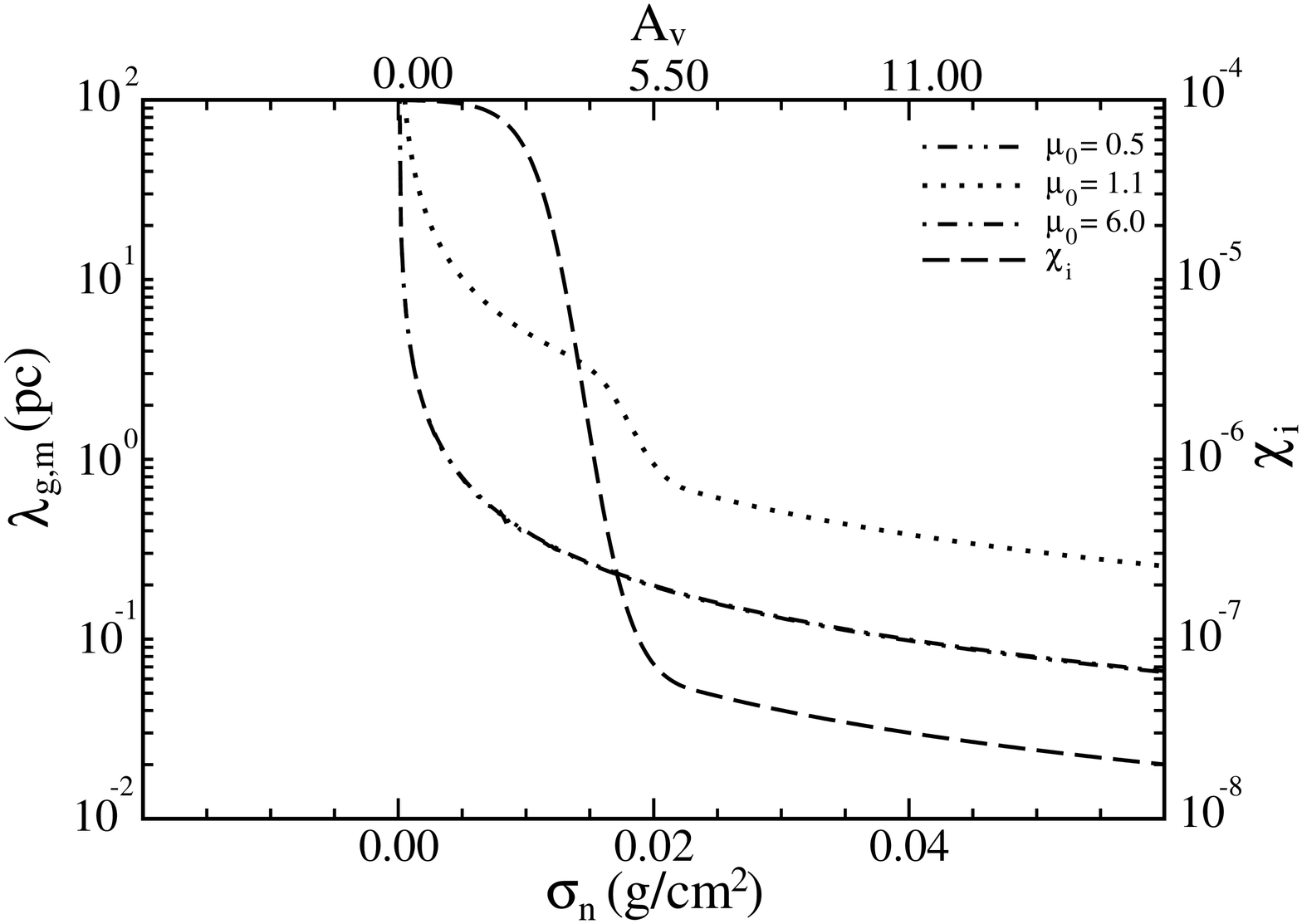}
\caption{Derived values of minimum growth time scale ($\tau_{g,m}$, top) and corresponding length scale $\lambda_{g,m}$, bottom) as a function of neutral column density. Plots show curves for assumed mass-to-flux ratios $\mu_{0} = 0.5$ (dash-dot-dot line), $\mu_{0} = 1.1$ (dotted line) and $\mu_{0} = 6.0$ (dash-dot line). Also shown is the ionization curve from Figure~\ref{xivssigma} (dashed line).}
\label{tlvssigma}
\end{figure}
\end{center}

For each value of $\tau_{ni,0}$, assuming a constant $\mu_{0}$, $\tau_{g,m}$ and $\lambda_{g,m}$ can be determined in a similar manner as in Figure~\ref{fig:vsmu}. Figure~\ref{tlvssigma} shows the minimum time scale (top) and corresponding length scale (bottom) as a function of $\sigma_{n}$ for assumed mass-to-flux ratios of 0.5 (dash-dot-dot line), 1.1 (dotted line) and 6.0 (dash-dotted line). Also plotted for reference is the ionization curve from Figure~\ref{xivssigma} (dashed line). Note that in the lower panel, the derived values for the $\mu_{0} = 0.5$ and $\mu_{0} = 6.0$ curves are almost equal and lie nearly on top of one another. As depictedin the right hand panel of Figure~\ref{fig:vsmu}, this correspondence of values is evident in the locations of the curves at these values of the mass-to-flux ratio.   
\begin{center}
\begin{table*}
\centering
\caption{Derived time and corresponding length scales for visual extinction thresholds}
\begin{tabular}{cccccccccc}
\hline
&& \multicolumn{2}{c}{$A_{v} = 3$} && \multicolumn{2}{c}{$A_{v} = 5$} && \multicolumn{2}{c}{$A_{v} = 8$}\\
\cline{3-4}\cline{6-7}\cline{9-10}
$\mu_{0}$ && $\lambda_{g,m}$ (pc) & $\tau_{g}$ (Myr)  && $\lambda_{g,m}$ (pc)& $\tau_{g}$ (Myr) && $\lambda_{g,m}$ (pc) & $\tau_{g}$ (Myr)\\
\hline
\hline
0.5 && 0.363 & 931  && 0.215  & 3.07 && 0.134 & 0.997 \\
1.1 && 4.717 & 2.44 && 1.520  & 1.09 && 0.883 & 0.528 \\
2.0 && 0.650 & 0.44 && 0.372  & 0.26 && 0.224 & 0.157 \\
6.0 && 0.368 & 0.29 && 0.218  & 0.17 && 0.136 & 0.107 \\ 
\hline
\label{avcomp}
\end{tabular}
\end{table*}
\end{center}

For a diffuse molecular cloud, (i.e., $A_{v} < 3$), the mass-to-flux ratio can be expected to be either sub- or transcritical as discussed in Section~\ref{intro}. Figure~\ref{tlvssigma} shows that while the length scale for fragmentation of a subcritical region is small, the time scale is essentially infinite, therefore direct fragmentation of a diffuse subcritical cloud into cores is unlikely. Instead, the more likely scenario is the fragmentation of a transcritical cloud. Under this assumption, the length scale for fragmentation is on the order of several parsecs and the time scale for collapse is no longer effectively infinite (see Figure~\ref{tlvssigma}, dotted line) therefore clumps can form. As the newly formed clump starts to contract, the density increases, and consequently the visual extinction. This contraction will eventually push the clump past a critical threshold $A_{v,crit}$ where the ionization fraction, and the length and time scales for collapse decrease dramatically. When this occurs, high density regions within the clump may subfragment and start collapsing independently from the clump. The time scale for the collapse of these subfragments is dictated by the local mass-to-flux ratio. Trans- and supercritical subfragments will collapse faster than the parent clump while subcritical subfragments will not.

From these plots, relevant time and length scales corresponding to the collapse of clumps and cores can be determined. Table~\ref{avcomp} shows these scales for the three extinction thresholds and four values of $\mu_{0}$. For each extinction threshold, the variation of $\lambda_{g,m}$ and $\tau_{g,m}$ across the different values of $\mu_{0}$ further highlights the trends shown in Figure~\ref{fig:vsmu}. Looking at a specific $\mu_{0}$ value, the length and time scales for collapse become progressively smaller for increasing extinction. Of particular interest are the comparisons between the length and time scales for the formation of clumps versus cores. Assuming a clump has $\mu_{0} = 1.1$ and $A_{v} \sim 3$ and a subfragment has an $A_{v} \sim 5$, $\lambda_{\rm clump}/\lambda_{\rm core} = 3 - 22$ and $\tau_{\rm clump}/\tau_{\rm core} = 2.2 - 14$ for $\mu_{0} \ge 1.1$ in the core. For $\mu_{0} < 1$ in the core, $\tau_{\rm clump}/\tau_{\rm core} \le 1$ indicating that a subcritical subfragment will not collapse before the parent clump does.   

Note that these time scales represent one $e$-folding of the density. Depending on the initial density, it may take several $e$-folding times for a structure to reach the threshold density that defines it as a core. Taking all of this into account, this analysis suggests that a smaller, more dense region can form within a larger, more diffuse region and can collapse within the time scale of collapse for the larger region. The disparity of time and length scales for collapse between the clump and core is \textit{far greater} for a model with magnetic fields and ambipolar diffusion than it is for a simple hydrodynamic model.
 
\begin{table*}
\caption{Derived length scales for B1-E}
\centering
\begin{tabular}{cccccc}
\hline
&& \multicolumn{4}{c}{$\lambda_{g,m}$ (pc)}\\
\cline{3-6}
$A_{v}$ (mag) & $N_{n}$ (10$^{21}$ cm$^{-2}$)& $\mu_{0} = 0.5$ & $\mu_{0} = 1.1$ & $\mu_{0} = 2.0$ &  $\mu_{0} = 6.0$ \\
\hline
\hline
2.9  & 2.7 & 0.373 & 4.85 & 0.668 & 0.378\\
6.8  & 6.3 & 0.160 & 1.05 & 0.268 & 0.162\\
10.6 & 9.9 & 0.102 & 0.66 & 0.170 & 0.103\\
\hline
\label{clumpcomp}
\end{tabular}
\end{table*}
\begin{table*}
\centering
\caption{Mean derived length scales for fragments within B1-E}
\begin{tabular}{ccccccc}
\hline
&&& \multicolumn{4}{c}{$\lambda_{g,m}$ (pc)}\\
\cline{4-7}
Object & $N_{n}$ (10$^{21}$ cm$^{-2}$)& $D_{obs}$ (pc) &$\mu_{0} = 0.5$ & $\mu_{0} = 1.1$ & $\mu_{0} = 2.0$ & $\mu_{0} = 6.0$ \\
\hline
\hline
B1-E1 & 12.1 - 35.0 & 0.066 & 0.056 & 0.367 & 0.093 & 0.056\\
B1-E2 & 13.4 - 30.0 & 0.064 & 0.054 & 0.357 & 0.091 & 0.055\\
B1-E3 & 8.9 - 26.8  & 0.072 & 0.075 & 0.493 & 0.126 & 0.076\\
B1-E4 & 7.5 - 31.3  & 0.056 & 0.083 & 0.547 & 0.139 & 0.084\\
B1-E5 & 6.5 - 24.5 & 0.087 & 0.098 & 0.645 & 0.164 & 0.100\\
B1-E6 & 9.8 - 29.1 & 0.055 & 0.069 & 0.453 & 0.115 & 0.070\\
B1-E7 & 9.6 - 29.1 & 0.055 & 0.070 & 0.459 & 0.117 & 0.071\\
B1-E8 & 6.5 - 25.1 & 0.068 & 0.098 & 0.643 & 0.164 & 0.099\\
B1-E9 & 9.1 - 28.2 & 0.051 & 0.073 & 0.481 & 0.123 & 0.074\\
\hline
\label{subclumpcompavg}
\end{tabular}
\end{table*}

\subsection{Comparison to Observations}
\label{comptoobs}

The above analysis reveals several physical conditions that are important for the collapse of clumps and subclumps into cores. First, for regions with transcritical mass-to-flux ratios, the preferred length scale decreases as the initial column density/visual extinction of the region increases. Based on our ionization model (see Figure~\ref{xivssigma}), less dense regions ($A_{v}< 3-5$) have larger ionization fractions (smaller neutral-ion collision times) and thus yield a larger length scale for collapse while the denser regions ($A_{v} > 5$) have smaller ionization fraction (larger neutral-ion collision times) and thus yield much smaller collapse length scales. From a time scale point of view, it seems favorable for regions within a clump to fragment and start to collapse before the overall clump structure collapses or fragments. 

Observations of nearby molecular clouds, including (but not limited to) the Taurus molecular cloud \citep{Rebull2010,Hartmann2002,Schmalzl2010}, Rosette Molecular cloud \citep{WBS1995}, Pipe Nebula \citep{Frau2010,RZ2010} and Perseus molecular cloud \citep{Kirk2006,Sadavoy2012} show clumpy filamentary structure both with and without active star formation. In the following discussion we choose a small sample of clumps which show evidence of substructure, and perform in-depth case studies comparing our two-stage fragmentation model to the observed clump data. The three regions that we have chosen come from the Perseus Molecular cloud, the Taurus Molecular cloud, and the Pipe Nebula. These case studies are by no means a thorough compilation of all possible regions, but rather a representation of some interesting cases. This analysis can shed some insight into the possible magnetic field strengths within these structures. Note that the existence of subfragments within these chosen areas indicates that the clump has evolved to a point in the parameter space of Figure~\ref{tlvssigma} that is to the right of $A_{v,crit}$, although the initial fragmentation event that created the clumps would have occurred at lower densities on the left hand side of $A_{v,crit}$.  

\subsection{Case Study A: Perseus B1-E}
\label{b1e}

The Perseus molecular cloud is located at a distance of about 250 pc in the constellation Perseus. It shows clustered star formation primarily in two regions (NGC~1333 and IC~348) while elsewhere in the cloud it is relatively quiescent. A particularly interesting region is a $\sim0.1$ deg$^2$ clump roughly 0.7 degrees east of the B1 clump, coined B1-E \citep{Sadavoy2012}. This region is sandwiched between the two star forming regions (NGC~1333 to the west, and IC~348 to the east), however previous submillimeter observations of this region showed that despite its high extinction ($A_{v} > 5$), it contained neither dense cores nor young stellar objects \citep[][among others]{Kirk2006,Jorgensen2007}. However, recent data from the \textit{Herschel} Gould Belt survey and subsequent observations from the Green Bank Telescope (GBT) have revealed that there is substructure within this region \citep{Sadavoy2012}. In their paper, \citet{Sadavoy2012} suggest that the delay in the formation of cores within this region could be due to a strong magnetic field. In this case study, we use our linear analysis methods described above to see if theoretical models exhibiting the same density as the observations result in similar observed length scales. We also comment on the mass-to-flux ratio that is likely required to generate these observed length scales.

First, the B1-E clump as a whole was found to have a column density $N_{n} = (6.3 \pm 3.6)\times 10^{21} \rm ~cm^{-2}$ and a radius of 0.46 pc \citep{Sadavoy2012}. Taking into account the error range on the density, this corresponds to a visual extinction range of 2.89 to 10.59 magnitudes, with the mid-point value corresponding to a visual extinction of $\sim$~6.75. Table~\ref{clumpcomp} shows the derived length scales for collapse to be compared with the diameter of the structure, for the low, mid and high values of $A_{v}$ for $\mu_{0} = 0.5,~1.1,2.0$ and $6.0$. From this analysis, we see that one scenario stands out as a good fit. The diameter of the B1-E clump is $\sim$ 1 pc which matches the model with $A_{v} = 6.8$ and $\mu_{0} = 1.1$ ($\lambda_{g,m} = 1.05$ pc). All the other models can be ruled out because they result in length scales that are much smaller than the diameter of the region. The model with $A_{v} = 6.8$ and $\mu_{0} = 1.1$ corresponds to a growth time of about 0.5 Myr.

We now turn to the substructures identified within this clump. \citet{Sadavoy2012} list a range of densities for each fragment. Table~\ref{subclumpcompavg} shows the length scales, $\lambda_{g,m}$, for each fragment. We have calculated $\lambda_{g,m}$ for both the upper and lower limits of these densities. For clarity, we present only the mean value of $\lambda_{g,m}$ for each of the four mass-to-flux ratios. Looking at the derived values, we see that those found for both $\mu_{0}=0.5$ and $\mu_{0}=6.0$ are consistent with the observed diameter D$_{\rm obs}$ for each fragment. This suggests that the cores are in either a strongly or weakly magnetized region depending on if the mass-to-flux ratio is sub- or supercritical respectively. 

The lack of star formation in this region is evidence for either a slower fragmentation process or a later fragmentation event. The former possibility requires the presence of a strong magnetic field in the region. Formation of subcritical cores within a transcritical clump would require fragmentation along the magnetic field lines, which is not currently included in our model. \citet{Sadavoy2012} note that this region is likely isolated from the surrounding regions given it does not fall into the age gradient provided by the two closest regions (IC 348 and NCG 1333). This lends credence to the possibility that this region is younger than the surroundings. However, if the clump/core/star formation threshold values hold, $A_{v}$ values for the cores of $\sim$ 7-13 mag suggest that they should have formed stars. The lack of observed stars within this region therefore suggests that collapse is being slowed. \citet{Sadavoy2012} concluded that this delay is due to a strong magnetic field. 

Currently there are no magnetic field measurements specifically for the B1-E region, however measurements of the nearby B1 region have shown the existence of a strong magnetic field ($B = 19-27 \mu$G) \citep{Goodman1989,Crutcher1993}. The mass-to-flux ratio of a cloud can be written in terms of the density and field strength \citep{Chapman2011} as,
\begin{equation}
\mu = 7.6 N_{||}(H_{2})/B_{tot},
\end{equation}
where $N_{||}(H_{2})$ is the column density in units of 10$^{21}$ cm$^{-2}$ along a magnetic flux tube and $B_{tot}$ is the total magnetic field strength in $\mu$G. Observable quantities are the column density of molecular hydrogen, N(H$_{2}$), and the magnetic field component parallel to the line of sight ($B_{||})$. Projection effects between $N_{||}(H_{2})/B_{tot}$ and the observed quantity $N(H_{2})/B_{||}$ will overestimate $\mu$ by an average factor of 3 assuming a random orientation of the magnetic field with respect to the line of sight \citep{HC2005}. By combining this with the conversion factor for $H_{2}$ number density to visual extinction, namely $N(H_{2}) = 9.4 \times 10^{20} A_{v}$, the mass-to-flux ratio of a region can be calculated in terms of the visual extinction and line-of-sight component of the magnetic field,
\begin{equation}
\mu = 2.4~A_{v} /B_{||},
\label{obsmu}
\end{equation}
where $A_{v}$ is in mag and $B_{||}$ is in $\mu$G.

Assuming that the magnetic field in B1-E is similar to its neighboring region B1 and applying this equation to the values for the cores, we find that the majority of the mass-to-flux ratios fall into the transcritical to supercritical regimes (0.7-4.42). Some of the lower end densities for the cores give significantly subcritical values ($\mu \sim 0.32$), however given the range in the observed densities for each core, these lower values are likely not indicative of the true density. Less extreme mass-to-flux ratios (i.e., $\mu \sim 4$ vs $\mu \sim 6$) suggest that the cores may be larger than defined in \citet{Sadavoy2012}. Following this scenario, a strong magnetic field combined with higher densities results in regions that are transcritical which collapse on longer time scales than their supercritical counterparts.

Based on the above analysis, we conclude that this region is evolving at a slower pace than the neighbor-
\begin{table*}
\centering
\caption{Derived length scales for B218}
\begin{tabular}{cccccc}
\hline
&& \multicolumn{4}{c}{$\lambda_{g,m}$ (pc)}\\
\cline{3-6}
$A_{v}$ (mag)& $N_{n}$ (10$^{21}$ cm$^{-2}$)& $\mu_{0} = 0.5$ & $\mu_{0} = 1.1$ & $\mu_{0} = 2.0$ &  $\mu_{0} = 6.0$ \\
\hline
\hline
3 & 2.80 & 0.363 & 4.717 & 0.650 & 0.368 \\
5 & 4.67 & 0.215 & 1.520 & 0.372 & 0.218 \\
10 & 9.39 & 0.1074 & 0.7047 & 0.1795 & 0.1090 \\
\hline
\label{clumpcomptaurus}
\end{tabular}
\end{table*}
ing regions and will need to attain greater column densities than the typical threshold value ($A_{v} \sim 8$) to push the cores into the supercritical regime, due to the strong magnetic field present. Observations of the entire Perseus region by \citet{Goodman1990} qualitatively indicate that the magnetic field strengths in NGC 1333 and IC 348 are weaker than in the B1 and B1-E regions. Assuming these observations are fairly complete, the presence of a weaker field in the regions which exhibit star formation and a greater field in those regions which do not show signs of star formation are generally consistent.

\subsection{Case Study B: Taurus B218}
\label{b218}

\begin{table*}
\centering
\caption{Derived length scales for fragments within B218}
\begin{tabular}{ccccccc}
\hline
&&& \multicolumn{4}{c}{$\lambda_{g,m}$ (pc)}\\
\cline{4-7}
Object & $N_{n}$ (10$^{21}$ cm$^{-2}$)& $D_{obs}$ (pc) &$\mu_{0} = 0.5$ & $\mu_{0} = 1.1$ & $\mu_{0} = 2.0$ &  $\mu_{0} = 6.0$ \\
\hline
\hline
B218-1 & 10.4 & 0.26 & 0.097 & 0.633 & 0.162 & 0.098\\
B218-2 & 5.9  & 0.24 & 0.170 & 1.116 & 0.284 & 0.172\\
B218-3 & 10.9 & 0.16 & 0.093 & 0.609 & 0.155 & 0.094\\
B218-4 & 9.4  & 0.16 & 0.107 & 0.705 & 0.180 & 0.109\\
B218-5 & 6.5  & 0.14 & 0.156 & 1.020 & 0.260 & 0.158\\
B218-6 & 6.5  & 0.14 & 0.156 & 1.020 & 0.260 & 0.158\\
\hline
\label{subclumpcomptaurus}
\end{tabular}
\end{table*}
The Taurus molecular cloud is one of the closest star forming regions, at a distance of only 137$\pm$10 pc \citep{Torres2007}. It is known to be host to at least 250 young stellar objects \citep{Rebull2010} that are distributed along the filamentary structures within the cloud \citep{Hartmann2002}. \citet{Schmalzl2010} present a high resolution column density map of the L1495 filament. Extinction maps of this region show that the lowest visual extinction is on the order of $A_{v}$ = 5 mag and numerous dense fragments exist with peak extinction values of $A_{v} \geq 15$ mag. Specifically, there are 4 fragments that coincide with fairly large dense core populations. These fragments are four known Barnard Objects (B213, B216, B217 and B218) containing 14, 10, 8 and 6 dense cores respectively. The size of the Barnard Objects are $\sim$~1~pc across while the radii of the cores are $\sim$~0.1~pc. For our analysis, we choose to focus on B218. 

B218 is an elongated object with an area of 0.13 pc$^{2}$. The major axis is 0.82 pc while the minor axis is 0.35 pc. \citet{Schmalzl2010} quote a surface density $N_{n} = 9.39\times 10^{21}$ cm$^{-2}$ for this region, corresponding to a visual extinction of about 10 mag. However, this only includes regions that have $A_{v} \ge 5$ mag. Other observations of the region suggest an overall visual extinction of the B218 object to be about 3 mag \citep{Gaida1984}. For our analysis, we will look at three extinction values, namely $A_{v} = 3.0, 5.0$, and $10$. Table~\ref{clumpcomptaurus} shows the derived length scales $\lambda_{g,m}$ for $\mu_{0} = 0.5,~1.1,2.0$, and $6.0$. From this analysis, assuming that the length of the major axis corresponds to the collapse length, we see there is no exact match between the derived and observed length scale for the extinction values used. However, based upon the size scales derived for the various mass-to-flux ratios, we can conclude that the clump is transcritical; the length scales for the other mass-to-flux ratios are too small under all $A_{v}$ considerations. Specifically, the observed length scale falls within the parameter space defined by $5$ mag $\le A_{v} < 10$ mag and $0.9~ \ltsimeq ~\mu_{0}~ \ltsimeq ~1.1$. A better match could be found by either increasing the visual extinction or decreasing value of the mass-to-flux ratio. Given that the mass-to-flux ratio will increase as the density increases, we suggest the former possibility; the visual extinction of the region is likely between 5 and 10 mag. Looking back at Table~\ref{avcomp}, we see that a region with an $A_{v} = 8$ and $\mu_{0} = 1.1$ gives a length scale more on par with the major axis of B218. 

From this analysis we can see that by identifying the major axis as the collapse length, the mass-to-flux ratio of the clump is estimated to be $0.9~\ltsimeq ~\mu_{0} ~\ltsimeq ~1.1$. Recent polarization observations of the Taurus region that estimate the magnetic field strength based on the dispersion of polarization directions \citep{Chapman2011} show that the entire cloud complex which contains B213, B216, B217 and B218 (defined collectively in their paper as B213) is significantly subcritical. These two results are in direct conflict with each other. The main reason for this discrepancy is the lumping together of the four Barnard objects into one larger region. Specifically there is a discrepancy in the visual extinction of the regions. \citet{Chapman2011} cite a visual extinction for their B213 region of $\sim 2$ mag while all four of the Barnard objects that make up this region are quoted to have visual extinction values on the order of 10 mag \citep{Schmalzl2010}. However, based on the analysis above, we see that the actual visual extinction of the B218 region is implied to be about 8 mag. Using this value of $A_{v}$ and the magnetic field strengths of B213 measured by \citet{Chapman2011} in Equation~(\ref{obsmu}) results in only somewhat subcritical values  ($\mu \sim 0.7$) as compared to those computed using $A_{v} = 2$. Also, given that the region defined as B213 in \citet{Chapman2011} covers a much larger area than the B218 region, it is plausible that the local magnetic field strength in B218 could be greater than the global field of their B213 region. Taking errors into account, it is possible for these regions to have mass-to-flux ratios that are closer to the critical value, as suggested by our analysis. 

Turning to the substructures, \citet{Schmalzl2010} list six high density regions within B218. Table~\ref{subclumpcomptaurus} shows the derived length scales $\lambda_{g,m}$ for each fragment. In this case, there is no clear cut mass-to-flux ratio regime that yield a match between $\lambda_{g,m}$ and $D_{\rm obs}$ for all of the cores. Fragments B218 2-4 have diameters that agree well with $\lambda_{g,m}$ in the $\mu_{0} = 2.0$ regime while fragments 5 and 6 agree well with $\lambda_{g,m}$ in the $\mu_{0} = 0.5$ and $\mu_{0} = 6.0$ regimes. Finally, fragment 1 is consistent with a mildly supercritical mass-to-flux ratio ($1.1 < \mu_{0} < 2.0$). As with the B1-E region, the lack of observed star formation within B218 is likely due to the strong magnetic field. Overall, the 
\begin{table*}
\centering
\caption{Derived length scales for B59-09}
\begin{tabular}{ccccccc}
\hline
&&& \multicolumn{4}{c}{$\lambda_{g,m}$ (pc)}\\
\cline{4-7}
$A_{v}$ (mag)& $N_{n}$ (10$^{21}$ cm$^{-2}$) & $D_{obs}$ (pc) &$\mu_{0} = 0.5$ & $\mu_{0} = 1.1$ & $\mu_{0} = 2.0$ &  $\mu_{0} = 6.0$ \\
\hline
\hline
10   & 9.35 & 0.3  & 0.1078 & 0.7071 & 0.1802 & 0.1093 \\
31.6 & 29.6 & 0.11 & 0.0341 & 0.2233 & 0.0569 & 0.0345 \\ 
\hline
\label{b59}
\end{tabular}
\end{table*}
\begin{table*}
\centering
\caption{Derived length scales for select cores within the Pipe Nebula}
\begin{tabular}{ccccccc}
\hline
&&& \multicolumn{4}{c}{$\lambda_{g,m}$ (pc)}\\
\cline{4-7}
Object & $N_{n}$ (10$^{21}$ cm$^{-2}$) & $D_{obs}$ (pc) &$\mu_{0} = 0.5$ & $\mu_{0} = 1.1$ & $\mu_{0} = 2.0$ &  $\mu_{0} = 6.0$ \\
\hline
\hline
14 & 13.3 & 0.071  & 0.076 & 0.50 & 0.127 & 0.077\\
40 & 11.1 & 0.104  & 0.091 & 0.60 & 0.152 & 0.092\\
48 & 6.1  & 0.127  & 0.164 & 1.08 & 0.274 & 0.166\\
109 & 47.6& 0.063 & 0.021 & 0.14 & 0.035 & 0.021\\
\hline
\label{subclumpcomppipe}
\end{tabular}
\end{table*}
analysis of this region yields results that are consistent with the results of Section~\ref{modcalc}.

\subsection{Case Study C: Pipe Nebula}

Finally, the Pipe Nebula is slightly further away than the Taurus molecular cloud, at a distance of 145 pc. Unlike the previous two regions, the Pipe Nebula, although filamentary, shows little evidence of star formation; the only evidence being in B59. Recent studies \citep{Frau2010,RZ2010} show that there exists evolutionary variation across the nebula. \citet{Frau2010} propose that the fragmentation in the bowl corresponds to early stages of evolution, the collapsing material in the stem corresponds to an intermediate phase, and the star formation within B59 corresponds to the latest stage of evolution. This in itself makes the Pipe Nebula an interesting observational case on its own, as it seems to be providing us with several of the snap shots needed to decipher cloud to star evolution in one location. Dust emission maps show that there is a difference in the amount of structure within each region \citep{RZ2010} that follows the evolutionary gradient proposed by \citet{Frau2010}. The maps of the shank/stem regions show that they are much more diffuse than the B59 region, suggesting that they are indeed less evolved. \citet{RZ2010} find 220 extinction peaks within the stem, shank, bowl, and smoke regions of the nebula. All of the extinction peaks are within regions that have visual extinctions greater than the threshold value quoted by \citet{Ruffle1998}. In addition, they all have radii that are on the order of a tenth of a parsec or less, which is in line with the sizes found in our analysis. Analysis of the location of peaks within the various regions of the cloud show that there is a high density of peaks within the bowl and B59 \citep{RZ2010}. 

The previous two case studies have looked at the relationship between the length scales for clumps and cores in regions with little to no star formation. For this region, rather than repeating this type of analysis, we use our model to look at two different phenomena. First, we will apply our model to B59 as a whole. Second, we examine the four cores studied by \citet{Frau2010}. The four cores in question are core 14, 40, 48 and 109, which come from three distinct regions of the Pipe Nebula; 14 resides in B59, 40 and 48 reside in the stem and core 109 is in the bowl.

B59 is the only region within the Pipe Nebula that exhibits active star formation. It is itself split up into several clumps as shown by \citet[][Figure 6]{RZ2009}. Although all clumps show evidence of cores within them, only B59-09 harbours visible YSOs. As such, B59-09 provides the unique opportunity to look at all three levels of evolution within star formation: clump, cores and YSOs. Figure~2 of \citet{RZ2012} shows the centrally condensed $A_{v}$ profile of the B59-09ab cores, centered on the star forming region. Assuming a minimum visual extinction threshold for the region of 10 mag, we find that the B59-09 clump has a radius of $\sim$~30000~AU, which corresponds to a diameter $D_{obs}$~=~0.3~pc. Recent studies of B59 by \citet{RZ2012} show that the star forming core(s) within B59, specifically B59-09a and 09b, have a molecular hydrogen surface density of $N_{H_{2}}$~=~2.96$\times 10^{22}$~cm$^{-2}$ and a diameter $D_{obs}$~=~0.11~pc. This surface density corresponds to a visual extinction of 31.6. 

Table~\ref{b59} shows the results of our analysis for both the extended and local clump regions of B59-09, with $A_{v} = 10$ and 31.6 respectively. Focussing on the extended region first and comparing the derived length scales to the observed, we conclude that the mass-to-flux ratio of the region is somewhere between 1.1 and 2.0. This means that the star forming clump B59-09 as a whole is moving out of the transcritical regime and is becoming supercritical. Diving a little deeper into this region, the second entry in Table~\ref{b59} shows that the high density region which contains the YSOs, if considered separately, is also globally between $\mu_{0} = 1.1$ and $\mu_{0} = 2.0$. 

Going one step deeper, if the spacing between the YSOs ($\sim0.06$ pc) is assumed to have a near one-to-one relationship with the scale of the earlier fragmentation, our analysis can shed some light on the probable conditions in the region at that time. Table~\ref{b59} shows that for both $A_{v}$ values, in order to match the YSO spacing, the region is either significantly subcritical or supercritical. Evidence of YSOs within the region suggests the latter as the former have longer collapse times. 

With that in mind, our analysis implies that in order for the YSOs to form with such spacing, the local mass-to-flux ratio in the locations of current YSOs must have been larger than the surrounding region to promote collapse. Observations show that the magnetic field strength in B59 is 17$\mu$G \citep{Alves2008}. We can therefore narrow the visual extinction range in which the YSO fragmentation occurred. In order for a subregion to collapse, the mass-to-flux ratio must be sufficiently different from the background value. If we assume a minimum mass-to-flux ratio $\mu = 2.0$, for $B = 17$ $\mu$G, Equation~(\ref{obsmu}) yields a minimum visual extinction of 14 mag. Therefore, for the extinction range, 14~mag~$<~A_{v}~<$~31.6~mag, the observed magnetic field results in a corresponding mass-to-flux ratio range for the cores to be $2.0~<~\mu~<~4.5$.

We now consider the four cores from the various regions in the Pipe Nebula. Table~\ref{subclumpcomppipe} shows the results of this analysis. Comparing the derived values of $\lambda_{g,m}$ with $D_{\rm obs}$, cores 14, 40 and 48 seem to agree well with those in either the subcritical regime or highly supercritical regime ($\mu_{0} = 6.0$). Core 109 does not agree well with any of the three regimes, however it can be concluded that it is not extremely sub- or supercritical but rather somewhere between $\mu_{0}=$1.1 and 2.0. With regards to the evolutionary gradient proposed by \citet{Frau2010}, our length scales do show a trend to smaller length scales for cores 48, 40 and 14. Core 109 does not seem to fit in with this trend at all and given its size compared to core 40 could be mistaken as a star forming core. Comparison between the length scales for B59-09 and core 14 is not applicable since  the latter is not contained in B59-09.

The evolutionary gradient between B59 and the bowl of the Pipe Nebula is a direct consequence of the different magnetic field strengths present in the different regions; $B = 17, 30$, and 65 $\mu$G in B59, the stem, and the bowl respectively \citep{Alves2008}. The region with the lowest magnetic field has formed stars while the other two regions have only formed cores. This is consistent with the results of our other two case studies (see Sections~\ref{b1e} and \ref{b218}), whereby the delay in star formation is due to a strong magnetic field which forces the region to attain a higher column density before it becomes supercritical and can collapse. 

The current state of B59 (i.e., shows evidence of several cores/YSOs yet has not collapsed itself) is supported by its apparent transcritical nature. The age of the stellar cluster within B59-09 indicates that the clump has survived for longer than 10 free fall times \citep{RZ2012}. This again suggests that the magnetic field within the region is strong enough to delay the overall collapse of the B59-09 clump \citep{RZ2012} while the inner regions have attained sufficient conditions to form stars. Based on the current data, \citet{RZ2012} propose that B59-09 is currently  contracting toward a denser configuration. Our model of a transcritical clump is consistent with this assertion. 

\begin{center}
\begin{figure}
\includegraphics[width=0.35\textwidth,angle=-90]{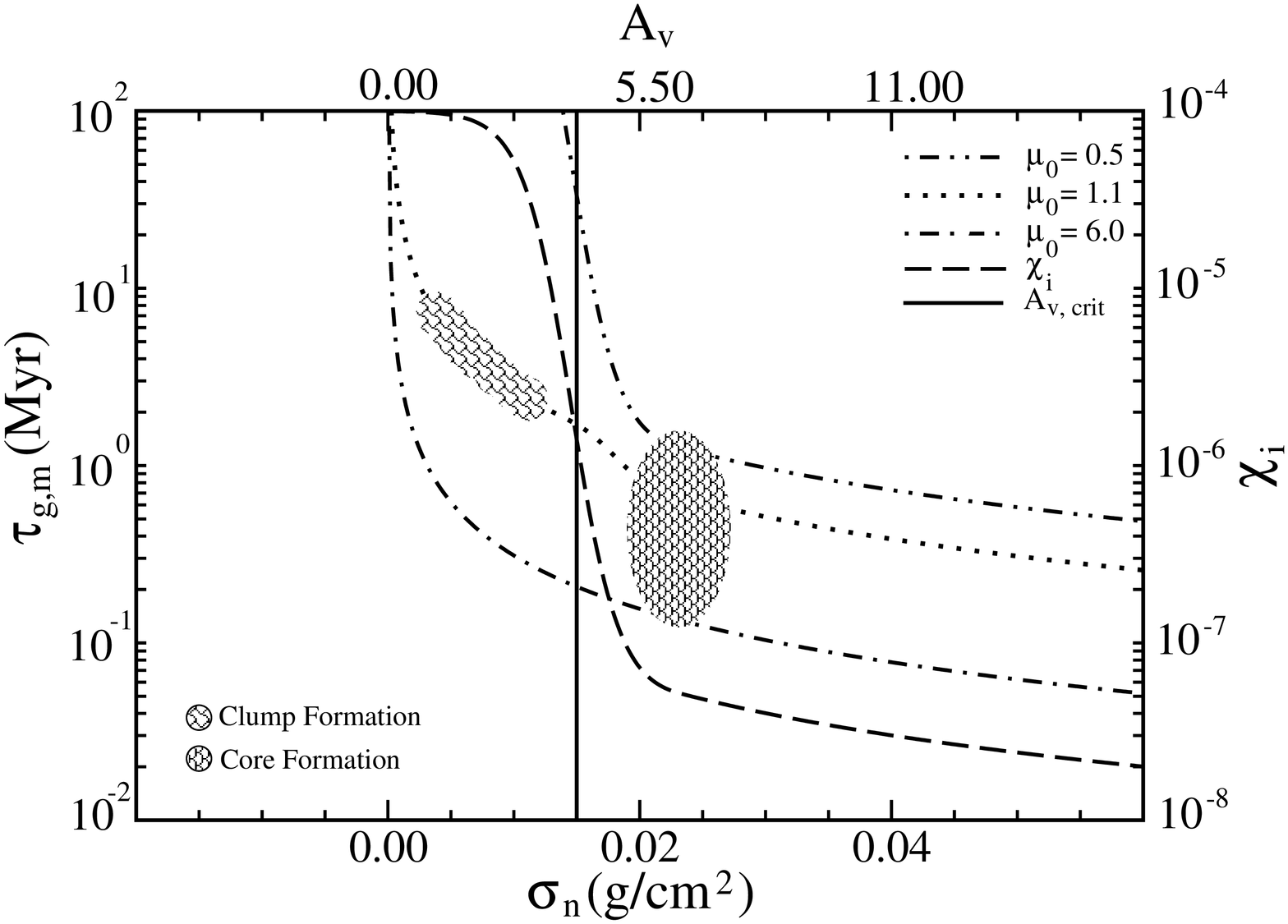}\\
\includegraphics[width=0.35\textwidth,angle=-90]{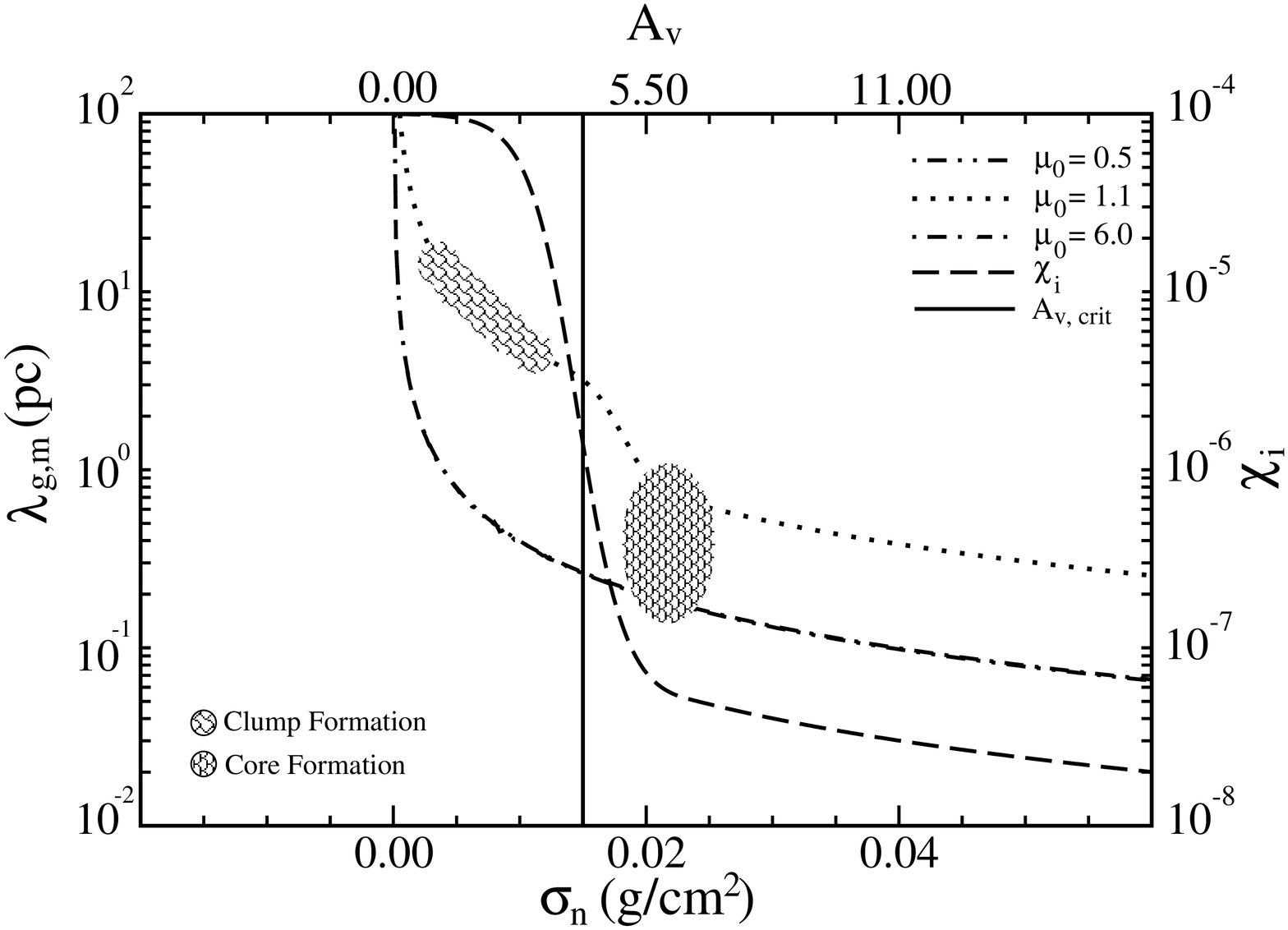}
\caption{Illustrative plots for two stage fragmentation scenario. Panels show the same mass-to-flux ratio curves as Figure~\ref{tlvssigma}. Shaded regions show the parameter space for the formation of clumps and cores respectively. }
\label{tlvssigma_enh}
\end{figure}
\end{center}

\section{Discussion}

The underlying model and subsequent results for the two-stage fragmentation scenario, as outlined in Section~\ref{modcalc}, propose various constraints on the conditions required for such a model to work in nature. Namely, the clump must form in an initially transcritical diffuse environment while core formation requires a denser supercritical region within the parent clump. From the three case studies presented above, we have found that the observed length scales and corresponding mass-to-flux ratios determined from our analysis tend to support this scenario. In general it seems that parent clumps exhibit mass-to-flux ratios that are transcritical ($0.9 < \mu_{0} < 1.1$). The column densities of these regions place them to the right hand side of $A_{v,crit}$ (see Figure~\ref{tlvssigma}). This is consistent with our model since formation of subfragments requires that the clump has already crossed over the critical $A_{v}$ threshold.

Our analysis also shows that the fragments within these clumps exhibit length scales that are indicative of regions that are either very subcritical or very supercritical. In order to truly delineate between these two extremes, high resolution magnetic field data is required, however without such data, some conjectures about the region can still be drawn. As shown by e.g., \citet{Ciolek1994}, the mass-to-flux ratio increases as the density of the region increases. The denser subfragments within the clump would be expected to have a greater mass-to-flux ratio than the global clump, allowing it to collapse faster than the clump itself (see Figure~\ref{tlvssigma}). It is possible for a minority of subfragments to locally have a subcritical mass-to-flux ratio, however these subfragments would likely not be able to collapse faster than their surroundings due to the long time scale required. Although the lack of magnetic field data for  most of the regions makes definitively choosing one mass-to-flux regime over the other impossible, it can be assumed that regions with active star formation must be within locally supercritical subfragments, while regions with little to no star-formation are either too young for the subfragments to have collapsed or are largely subcritical. 

Given the variability of the conditions within cloud regions, Figure~\ref{tlvssigma_enh} shows an enhanced picture of the two-stage fragmentation model. The bottom panel shows the fragmentation length scale as a function of column density. In the diffuse regions of the clouds, as indicated earlier, we expect the mass-to-flux ratio be in the transcritical regime. The parameter space where we expect the fragmentation of the gas into parsec size clumps is indicated by the hatched region to the left of $A_{v,\rm crit}$. As shown in the upper panel, this hatched region corresponds to fragmentation time scales on the order of 2-10 Myr. As the density of the region increases, as described in Section~\ref{modcalc}, the region will cross over to the right hand side of $A_{v,\rm crit}$. The hatched region on this side of $A_{v,\rm crit}$ represents the parameter space for subfragmentation within the clump. As shown, a wide range of mass-to-flux ratios are considered able to fragment based on the decrease of the fragmentation length scale. However, the equivalent region in the upper panel of Figure~\ref{tlvssigma_enh} shows that although regions within this range of mass-to-flux ratios can fragment, only the ones with trans- or supercritical mass-to-flux ratios (dotted and dash-dot lines respectively) are able to collapse at a rate that is faster than that of the transcritical clump itself.

\section{Summary}

We have studied the effect of ambipolar diffusion on the formation of stellar clusters using the results of linear analysis. By combining the linear analysis with realistic ionization profiles for a molecular cloud, our analysis has yielded a two-stage fragmentation model for clustered star formation that includes the formation of clumps and their subsequent subfragmentation. We present several interesting concepts that are worth noting. 

\begin{enumerate}
\item Linear analysis shows that there are varying length and time scales for collapse depending on both the mass-to-flux ratio and the neutral-ion collision time of the region \citep[][this work]{mor91,CB2006}. Transcritical mass-to-flux ratios can have significantly larger fragmentation scales compared to super- and subcritical regions (Figure~\ref{fig:vsmu}; right). Subcritical cores have significantly longer collapse time scales compared to their trans- and supercritical counterparts (Figure~\ref{fig:vsmu}, left). Increased neutral-ion collision times serve to decrease the length and time scales for collapse, particularly for the transcritical and subcritical regimes.

\item A combination of the linear analysis and ionization profile shows that molecular cloud conditions are favorable for a two-stage fragmentation process for the formation of stars. This model suggests that an initially diffuse, transcritical (i.e., $\mu_{0}$ is approximately unity) cloud can undergo an initial fragmentation on parsec scales and then undergo a second fragmentation event. This second event occurs once the density of the clump increases past the threshold where the ionization fraction drops dramatically and the length scale for fragmentation decreases steeply (Figure~\ref{tlvssigma}).

\item Comparison with several observed core and star forming regions show that the clump sizes are consistent with their mass-to-flux ratio being transcritical, while the core length scales imply they could be sub- or supercritical. Lack of star formation within the observed regions (excluding B59) could suggest subcritical cores. However, analysis of the measured magnetic field for Perseus and Taurus showed that these regions are at least mildly supercritical.

\item The threshold values for clump/core/star formation as outlined in Section~\ref{ctoc} are dependent on the region in question and should not be used as strict values. As shown in our case studies, regions with higher intrinsic magnetic fields will require a greater column density (higher visual extinction) to push the region to a transcritical and supercritical mass-to-flux ratio. The Pipe Nebula is a perfect example of this, whereby cores in all three regions (B59, stem and bowl) have densities that exceed the star formation threshold $A_{v} \sim 8$ in other clouds, but only the region with the lowest magnetic field strength (B59) currently shows evidence of young stellar objects. 
\end{enumerate}

\section*{Acknowledgments}
We thank Telemachos Mouschovias for insightful comments on the manuscript. NDB was supported by a scholarship from the Natural Science and Engineering Research Council (NSERC) of Canada. SB was supported by a Discovery Grant from NSERC.


\end{document}